\documentclass{jpp}
\usepackage{amsmath}
\usepackage{amssymb}	% Math symbols such as \mathbb
\usepackage{bbold}
\usepackage{bm} %allows msans to be used (for matrices' lettering)
\usepackage{cancel} % cancel terms out of an equation with \cancel{} or \cancelto{}{}
\usepackage{xcolor}
\usepackage[normalem]{ulem} %allows \sout{} for strikethrough characters
\usepackage{changepage}
\usepackage{float}
\usepackage{mathrsfs}
\usepackage{url}
\usepackage{xfrac}
\usepackage{graphicx}
\usepackage{lipsum} %Lorem ipsum
\graphicspath{{/}}
\newcommand{\comment}[1]{\ignorespaces} %for inline commenting
\renewcommand{\v}[1]{\ensuremath{\mathbf{#1}}} % for vectors
 
\renewcommand{\d}[2]{\frac{d #1}{d #2}} % for derivatives
\newcommand{\pd}[2]{\frac{\partial #1}{\partial #2}} 

\DeclareMathAlphabet{\mathsfit}{\encodingdefault}{\sfdefault}{m}{n}
\SetMathAlphabet{\mathsfit}{bold}{\encodingdefault}{\sfdefault}{bx}{n}
\newcommand{\ms}[1]{\bm{\mathsfit{#1}}} %block sans serif letters for matrices
\newcommand{\abs}[1]{\left| #1 \right|} % for absolute value
\newcommand{\ws}[1]{\mathsf{#1}} %For block letters in math mode

\makeatletter
\newcommand*\bigcdot{\mathpalette\bigcdot@{.5}}
\newcommand*\bigcdot@[2]{\mathbin{\vcenter{\hbox{\scalebox{#2}{$\m@th#1\bullet$}}}}}
\makeatother

\usepackage{accents}

\makeatletter %makes "@" just a letter
\renewcommand*\env@matrix[1][*\c@MaxMatrixCols c]{%
 \hskip -\arraycolsep
 \let\@ifnextchar\new@ifnextchar
 \array{#1}}
\makeatother

\newcommand*\mF{\mathcal{F}}

\newcommand*\mL{\mathcal{L}}

\newcommand*\mOne{\mathbb{1}}

\newcommand*\vphi{\varphi}

\newcommand*\mZ{\mathbb{Z}}

\newcommand*\mR{\mathbb{R}}

\newcommand*\md{\mathrm{d}}

\newcommand*\mf{\mathfrak{f}}
\newcommand*\mfd{\mathfrak{f}^*}
\newcommand*\mg{\mathfrak{g}}
\newcommand*\mgd{\mathfrak{g}^*}
\newcommand*\mgdd{\mathfrak{g}^{**}}

\let\emptyset\varnothing
\DeclareMathAlphabet{\mathpzc}{OT1}{pzc}{m}{it}

\DeclareFontFamily{U}{MnSymbolC}{}
\DeclareSymbolFont{MnSyC}{U}{MnSymbolC}{m}{n}
\DeclareFontShape{U}{MnSymbolC}{m}{n}{
 <-6> MnSymbolC5
 <6-7> MnSymbolC6
 <7-8> MnSymbolC7
 <8-9> MnSymbolC8
 <9-10> MnSymbolC9
 <10-12> MnSymbolC10
 <12-> MnSymbolC12}{}
\DeclareMathSymbol{\intprod}{\mathbin}{MnSyC}{'270}

\let\emptyset\varnothing

\usepackage{bbding}
\DeclareFontFamily{U}{MnSymbolC}{}
\DeclareSymbolFont{MnSyC}{U}{MnSymbolC}{m}{n}
\DeclareMathSymbol{\diamondplus}{\mathbin}{MnSyC}{"7C}
\DeclareMathSymbol{\diamonddot}{\mathbin}{MnSyC}{"7E}
\DeclareFontShape{U}{MnSymbolC}{m}{n}{
 <-6> MnSymbolC5
 <6-7> MnSymbolC6
 <7-8> MnSymbolC7
 <8-9> MnSymbolC8
 <9-10> MnSymbolC9
 <10-12> MnSymbolC10
 <12-> MnSymbolC12}{}

\makeatletter
\newcommand*{\defeq}{\mathrel{\rlap{%
 \raisebox{0.3ex}{$\m@th\cdot$}}%
 \raisebox{-0.3ex}{$\m@th\cdot$}}%
 =}
\newcommand*{\eqdef}{=\mathrel{\rlap{%
 \raisebox{0.3ex}{$\m@th\cdot$}}%
 \raisebox{-0.3ex}{$\m@th\cdot$}}%
 }
\makeatother

\usepackage{mathtools}

\usepackage{stmaryrd}

\usepackage{enumitem}

\definecolor{light-gray}{gray}{.85}
\newsavebox{\songboxbox}

\DeclareFontFamily{U}{matha}{\hyphenchar\font45}
\DeclareFontShape{U}{matha}{m}{n}{
 <5> <6> <7> <8> <9> <10> gen * matha
 <10.95> matha10 <12> <14.4> <17.28> <20.74> <24.88> matha12
 }{}
\makeatletter
\newcommand{\blandor}[1]{\mathbin{\@blandor{#1}}}
\newcommand{\@blandor}[1]{\mathchoice
 {\@@blandor{#1}{\tf@size}}
 {\@@blandor{#1}{\tf@size}}
 {\@@blandor{#1}{\sf@size}}
 {\@@blandor{#1}{\ssf@size}}
}
\newcommand{\@@blandor}[2]{%
 \raisebox{.1ex}{\rotatebox[origin=c]{#1}{%
 \fontsize{#2}{#2}\usefont{U}{matha}{m}{n}\symbol{\string"CE}}}%
}
\makeatother

\usepackage{slashed} %Feynman
\newcommand{\cmmnt}[1]{\ignorespaces} %for inline commenting
\newcommand{\sm}[2]{\left[\begin{smallmatrix}#1\\#2\end{smallmatrix}\right]} %for small matrices

\usepackage{environ}
\NewEnviron{eqn}{
\begin{align}\begin{split}
\BODY
\end{split}\end{align}
}

\usepackage{textgreek} %Allows writing \textdelta (e.g.) in text mode for biblioraphy

\makeatletter
\@addtoreset{footnote}{page}
\renewcommand{\thefootnote}{\ifcase\value{footnote}\or(*)\or
(**)\or(***)\or(****)\or(\#)\or(\#\#)\or(\#\#\#)\or(\#\#\#\#)\or($\infty$)\fi}
\makeatother
\usepackage[hang,multiple]{footmisc}

\begin{document}

%\preprint{}

\title{The geometric theory of charge conservation in particle-in-cell simulations} %Title of paper



\author{Alexander S. Glasser\aff{1,2}
 \corresp{\email{asg5@princeton.edu}},
 \and Hong Qin\aff{1,2}}

\affiliation{
\aff{1}Princeton Plasma Physics Laboratory, Princeton University, Princeton, New Jersey 08543
\aff{2}Department of Astrophysical Sciences, Princeton University, Princeton, New Jersey 08544
}


%\date{\today}

\maketitle %\maketitle must follow title, authors, abstract and \pacs


\begin{abstract}
In recent years, several gauge-symmetric particle-in-cell (PIC) methods have been developed whose simulations of particles and electromagnetic fields exactly conserve charge. While it is rightly observed that these methods' gauge symmetry gives rise to their charge conservation, this causal relationship has generally been asserted via \emph{ad hoc} derivations of the associated conservation laws. In this work, we develop a comprehensive theoretical grounding for charge conservation in gauge-symmetric Lagrangian and Hamiltonian PIC algorithms. For Lagrangian variational PIC methods, we apply Noether's second theorem to demonstrate that gauge symmetry gives rise to a local charge conservation law as an off-shell identity. For Hamiltonian splitting methods, we show that the momentum map establishes their charge conservation laws. We define a new class of algorithms---\emph{gauge-compatible splitting methods}---that exactly preserve the momentum map associated with a Hamiltonian system's gauge symmetry---even after time discretization. This class of algorithms affords splitting schemes a decided advantage over alternative Hamiltonian integrators. We apply this general technique to design a novel, explicit, symplectic, gauge-compatible splitting PIC method, whose momentum map yields an exact local charge conservation law. Our study clarifies the appropriate initial conditions for such schemes and examines their symplectic reduction.
\end{abstract}

\section{Introduction}
Particle-in-cell (PIC) methods have long been an indispensable tool in studies of theoretical plasma physics, with many algorithmic efforts tailored toward specific applications \citep{dawson_particle_1983,hockney_computer_1988,birdsall_plasma_1991,okuda_nonphysical_1972,cohen_implicit_1982,langdon_direct_1983,lee_gyrokinetic_1983,cohen_performance_1989,liewer_general_1989,friedman_multi-scale_1991,eastwood_virtual_1991,cary_explicit_1993,parker_gyrokinetic_1993,grote_new_1998,decyk_skeleton_1995,qin_three-dimensional_2000,qin_3d_2000,qiang_object-oriented_2000,chen_f_2003,qin_3d_2001,vay_mesh_2002,nieter_vorpal:_2004,huang_quickpic:_2006}. The literature counts several examples, in particular, of PIC methods that have been engineered to exactly conserve charge---to machine precision---by the use of various sophisticated numerical techniques \citep{villasenor_rigorous_1992,esirkepov_exact_2001,chen_energy-_2011,pukhov_particle--cell_2016}.

In recent years, elegant PIC methods have been developed that preserve the gauge symmetry of the plasmas they simulate. Such gauge-symmetric methods exactly conserve charge, not as the result of bespoke numerical methods, but as a natural consequence of preserving their systems' geometric structure. It was \citet{squire_geometric_2012} that first derived an exactly charge-conserving variational PIC scheme by imposing gauge symmetry on a discrete action. Several gauge-symmetric algorithms have since followed, especially in the form of Hamiltonian PIC schemes \citep{xiao_explicit_2015,he_hamiltonian_2015,he_hamiltonian_2016,qin_canonical_2016,kraus_gempic:_2017,xiao_structure-preserving_2018,xiao_field_2019}.

Many of these references note that the gauge symmetry of their algorithms guarantees exact charge conservation, but this fact is often unproven; the associated conservation laws are not always stated, let alone systematically derived. The absence of such derivations motivates a rigorous study of algorithmic conservation laws in PIC methods. In the present paper, we study Lagrangian variational and Hamiltonian splitting algorithms and derive their charge conservation laws from first principles. In so doing, we elucidate the requirements for gauge-symmetric codes to be charge conserving, and provide a general template for the derivation of conservation laws from the gauge symmetry of Lagrangian and Hamiltonian algorithms.

Our study of Hamiltonian systems, in particular, identifies a new and quite general class of algorithms---\emph{gauge-compatible splitting methods}---which guarantee the exact preservation of the momentum map associated with gauge symmetries in Hamiltonian systems---even after time discretization. We leverage this general classification in our present study and construct a novel gauge-compatible splitting PIC method. Our effort highlights the practical importance of solving for the momentum map in Hamiltonian algorithms, especially in determining the correct specification of their initial conditions.

This paper is presented in two parts (Sections~\ref{sectionN2T_Theory}-\ref{sectionN2T_Apply} and Sections~\ref{sectionSympRed}-\ref{sectionCanonCons_PIC}, respectively), each of which may be read independently. In Sections~\ref{sectionN2T_Theory}-\ref{sectionN2T_Apply}, we demonstrate the systematic derivation of an exact charge conservation law for the Lagrangian variational PIC method of \citet{squire_geometric_2012}. We discover this conservation law from the system's local gauge symmetry using Noether's second theorem (N2T) in a discrete setting, leveraging the formalism of \citet{hydon_extensions_2011}. Our effort draws upon the tools of discrete exterior calculus (DEC) \citep{hirani_discrete_2003,desbrun_discrete_2005}, and studies the subtleties involved in deriving conservation laws for degenerate Lagrangian systems.

In Sections~\ref{sectionSympRed}-\ref{sectionCanonCons_PIC}, we study the Hamiltonian formulation of the Vlasov-Maxwell system, its momentum map and its Poisson reduction \citep{marsden_reduction_1974,marsden_hamiltonian_1982,marsden_reduction_1986}. We provide an introduction \citep{souriau_structure_1970,marsden_introduction_1999} to the momentum map $\mu$ that arises from the gauge symmetry of the Vlasov-Maxwell system, and we demonstrate how ${\dot{\mu}=\{\mu,H\}=0}$ defines a continuous-time charge conservation law. We then define the class of gauge-compatible splitting methods, demonstrating their exact conservation laws in discrete time via the momentum map. In so doing, we highlight a significant advantage of such methods over alternative Hamiltonian integrators. We apply this general classification to design a new, explicit, symplectic, gauge-compatible splitting PIC algorithm for the Vlasov-Maxwell system, whose exact charge conservation law, initial conditions and symplectic reduction are systematically derived.

\section{A constructive review of Noether's second theorem\label{sectionN2T_Theory}}

Noether's first theorem (N1T) famously establishes a one-to-one correspondence between the symmetries of a Lagrangian and conservation laws satisfied by its Euler-Lagrange equations. However, in instances of degenerate Lagrangians (specifically, Lagrangians whose equations of motion are underdetermined) this correspondence, while true, is nevertheless weakened. In particular, in underdetermined systems there is no guarantee that \emph{non-trivial} symmetries are in one-to-one correspondence with \emph{non-trivial} conservation laws \citep{olver_applications_1986}. Such degenerate Lagrangians may be investigated using N2T, which describes the interdependence of equations of motion in Lagrangian systems with local gauge symmetry.

For present purposes, we regard a trivial conservation law as a conservation law that holds whether or not the equations of motion (EOM) are satisfied. Such a conservation law is said to hold \emph{off-shell}. (A dynamical field is said to be \emph{on-shell} when it obeys the equations of motion defining a system of interest; it is said to be \emph{off-shell} otherwise. A conservation law is said to hold on-shell if it is satisfied when restricted to on-shell fields; it is said to hold off-shell if satisfied even by off-shell fields.) In this way, trivial conservation laws are mathematical identities; they hold true regardless of any particular system dynamics.

N2T establishes a one-to-one correspondence between local gauge symmetries of a degenerate Lagrangian and off-shell differential identities of its Euler-Lagrange equations. Off-shell identities may at first appear to capture little information. Nevertheless, we will show that in variational PIC methods, the local charge conservation law ${\partial_t\rho+\nabla\cdot\v{J}=0}$ is just such an identity---a trivial conservation law that is independent of the dynamics of $\rho$ and $\v{J}$. Applying N2T, we will systematically derive this charge conservation law from the local gauge symmetry of a discrete Lagrangian.

N2T demonstrates that the redundancy of physical variables in a degenerate Lagrangian manifests in the interdependence of its EOM. In particular, N2T states that a general Lagrangian system admits a local gauge symmetry if and only if its EOM satisfy a differential identity of the form
\begin{eqn}
\mathcal{D}^1\ws{E}_{\alpha_1}(\mL)+\cdots+\mathcal{D}^q\ws{E}_{\alpha_q}(\mL)=0.
\label{generalDiffIdent}
\end{eqn}
Here, $\mathcal{D}^i$ represents an arbitrary differential operator (e.g., the Klein-Gordon operator: ${\mathcal{D}^i=\partial^2-m^2}$), and $\ws{E}_{\alpha_i}(\mL)$ denotes the Euler-Lagrange equation for the variable $\alpha_i$ (e.g., Maxwell's equation for $A_\nu$: ${\ws{E}_{A_\nu}(\mL)=\partial_\mu F^{\mu\nu}+J^\nu}$).

N1T can discover conservation laws that hold dynamically (on-shell), while N2T discovers differential identities that hold kinematically (off-shell). Although Eq.~(\ref{generalDiffIdent}) is an off-shell identity, it may nonetheless reveal valuable information for some Lagrangian systems. For discrete systems in particular, whose kinematics are sometimes less apparent or less studied, these differential identities can be especially enlightening.

In the following section, we briefly describe the formalism of \citet{hydon_extensions_2011}, which derives N2T's differential identities in the form of Eq.~(\ref{generalDiffIdent}) from the local gauge symmetries of a general Lagrangian system. As we shall see, this formalism is extensible to both continuous and discrete systems.

To begin, we recall the variation of an arbitrary action ${S=\int\md^4x\mL[\phi,\partial_\mu\phi,\dots]}$ for a field $\phi$ in flat space-time with coordinates ${x^\mu}$:
\begin{eqn}
\delta S&=\int\md^4x\left[\delta\phi\pd{\mL}{\phi}+\delta(\partial_\mu\phi)\pd{\mL}{(\partial_\mu\phi)}+\cdots\right]\\
&=\int\md^4x\left[\delta\phi\ws{E}_\phi(\mL)+\ws{D}_\mu\left(\delta\phi\pd{\mL}{(\partial_\mu\phi)}\right)+\cdots\right].
\label{DefineEInSitu}
\end{eqn}
In the above, we have employed the \emph{Euler operator} \citep{olver_applications_1986}
\begin{eqn}
\ws{E}_\phi&\defeq\sum\limits_J(-\ws{D})_J\pd{}{(\partial_J\phi)}\\
&=\left(\pd{}{\phi}-\ws{D}_\mu\pd{}{(\partial_\mu\phi)}+\ws{D}_{\mu\nu}\pd{}{(\partial_{\mu\nu}\phi)}-\cdots\right),
\label{EulerOperatorDef}
\end{eqn}
which discovers a Lagrangian's EOM by implementing a variational derivative with respect to a dynamical variable. The sum in Eq.~(\ref{EulerOperatorDef}) is taken over all \emph{multi-indices} $J$ of space-time variables---e.g., ${J\in\{\emptyset,x,tt,yzz,\dots\}}$---and ${\ws{D}_{\mu\nu}\equiv\ws{D}_\mu\ws{D}_\nu}$, where ${\ws{D}_\mu\equiv\d{}{x^\mu}=\partial_\mu+\phi_\mu\partial_\phi+\phi_{\mu\nu}\partial_{\phi_\nu}+\cdots}$ denotes a total derivative. (The notations ${\phi_\mu\equiv\partial_\mu\phi\equiv\partial\phi/\partial x^\mu,~\phi_{\mu\nu}
\equiv\partial_{\mu\nu}\phi\equiv\partial^2\phi/\partial x^\mu\partial x^\nu}$, etc. are to be used interchangeably.) For a Lagrangian with only first-order derivatives, the EOM of the field $\phi$ is thus given by its familiar form
\begin{eqn}
0=\ws{E}_\phi(\mL)=\pd{\mL}{\phi}-\ws{D}_\mu\left(\pd{\mL}{\phi_\mu}\right).
\end{eqn}

We now consider a Lagrangian $\mL[u^\alpha,u^\alpha_\mu,\dots]$ that depends on multiple fields $\{u^\alpha(x)\}$ and their derivatives. We suppose that ${S=\int\md^4x\mL}$ is invariant under an (infinite) group of local gauge transformations, each labelled by an arbitrary smooth function ${g(x)}$ over space-time. Such a gauge transformation may be envisioned as parametrizing a Lie group action on dynamical variables at each point of space-time individually, with the local transformation at each point determined by $g(x)$. (In the $U(1)$ gauge theory of electromagnetism, for example, the function ${g(x)=\theta(x)}$ may be associated with the local phase rotation of a matter field, ${\phi(x)\rightarrow e^{i\theta(x)}\phi}(x)$ at each point ${x\in\mR^4}$.)

We next define the infinitesimal generator $\v{v}_g$ of a gauge symmetry as a vector field on the product manifold ${X\times U=\{(x^\mu,u^\alpha)\}}$ (where $X$ represents space-time and $U$ the space of dynamical fields). Such a vector field may be realized as a differential operator:
\begin{eqn}
\v{v}_g&=\sum\limits_\alpha Q^\alpha[g]\partial_{u^\alpha}.
\label{defnOfVg}
\end{eqn}
Here, $Q^\alpha[g]$ are the so-called \emph{characteristics} of $\v{v}_g$, which generally depend on ${\{g(x),\partial_\mu g(x),\dots\}}$ and are defined for each dynamical variable $u^\alpha$. The symbol $\partial_{u^\alpha}$ defines a vector field on ${X\times U}$, which acts as a partial derivative with respect to $u^\alpha$ on functions of ${x^\mu}$ and ${u^\alpha}$. (We will clarify this with a concrete example momentarily.) We emphasize that the freedom to independently specify $g(x)$ at each point in space-time is what makes $\v{v}_g$ a local symmetry. A global symmetry, by contrast, would transform the fields at each point of space-time identically, such that ${g(x)=\text{const}}$.

Referring the reader to \citet{hydon_extensions_2011} for greater detail, we have now assembled the minimal formalism necessary to construct N2T's differential identity from a system's local gauge symmetry. Given an action ${S[u^\alpha,u^\alpha_\mu,\dots]=\int\md^4x\mL}$ that is invariant under the symmetry generator $\v{v}_g$ of Eq.~(\ref{defnOfVg}), N2T guarantees the following differential identity of its EOM:
\begin{eqn}
\ws{E}_g\left[\sum\limits_\alpha Q^\alpha[g]\ws{E}_{u^\alpha}(\mL)\right]=0.
\label{N2TDiffIdentity}
\end{eqn}
In this equation, $g(x)$ is treated as a dynamical variable, and its Euler operator $\ws{E}_g$ is applied to an expression involving each dynamical variable's EOM---${\ws{E}_{u^\alpha}(\mL)}$---and its corresponding characteristic in $\v{v}_g$---${Q^\alpha[g]}$.

Assuming that the characteristics ${Q^\alpha[g]}$ of $\v{v}_g$ are linear in $g$ and its derivatives, the final expression of Eq.~(\ref{N2TDiffIdentity}) is independent of $g$ (as we soon show by example), and correspondingly takes the form of Eq.~(\ref{generalDiffIdent}). Eq.~(\ref{N2TDiffIdentity}) is therefore an off-shell differential identity of the equations of motion; nowhere in this construction is the dynamical equation ${\ws{E}_{u^\alpha}(\mL)=0}$ enforced. Accordingly, using the characteristics of a Lagrangian system's local gauge symmetry, N2T's off-shell differential identity is easily discovered via Eq.~(\ref{N2TDiffIdentity}).

Before applying this method to the Vlasov-Maxwell system of interest in Section~\ref{sectionN2T_Apply}, we make the preceding N2T formalism more concrete with a brief example from the vacuum Maxwell action
\begin{eqn}
S=\int\md^4x\mL=-\frac{1}{4}\int\md^4xF_{\mu\nu}F^{\mu\nu},
\label{vacMaxAction}
\end{eqn}
where ${F_{\mu\nu}\equiv\partial_\mu A_\nu-\partial_\nu A_\mu}$. This action yields the familiar EOM
\begin{eqn}
0&=\ws{E}_{A_\sigma}(\mL)=\left[\pd{}{A_\sigma}-\ws{D}_\tau\pd{}{(\partial_\tau A_\sigma)}+\cdots\right]\mL=\partial_\tau F^{\tau\sigma}.
\label{vacMaxEOM}
\end{eqn}

We now observe that, for arbitrary smooth $\lambda(x)$, $S$ is invariant under the local gauge transformation ${A_\mu(x)\rightarrow A_\mu(x)-\partial_\mu\lambda(x)}$. The infinitesimal generator of this gauge transformation is given by the following vector field with characteristics ${Q^{A_\mu}[\lambda]=-\partial_\mu\lambda}$:
\begin{eqn}
\v{v}_\lambda&=-(\partial_\mu\lambda)\partial_{A_\mu}.
\label{defGaugeSymm}
\end{eqn}
Here, as above, the Einstein summation convention over $\mu$ is implicit. To see that this vector field is correct, note that the flow generated by $\v{v}_\lambda$ on the product manifold ${X\times\{A_\mu\}}$ transforms $A_\mu$ appropriately
\begin{eqn}
\exp\left[\v{v}_\lambda\right](x^\rho,A_\sigma)&=\left[\mOne+\v{v}_\lambda+\frac{1}{2!}\v{v}_\lambda^2+\cdots\right](x^\rho,A_\sigma)\\
&=\left[\mOne-(\partial_\mu\lambda)\partial_{A_\mu}+\frac{1}{2!}(\partial_\mu\lambda)^2\partial_{A_\mu}^2+\cdots\right](x^\rho,A_\sigma)\\
&=(x^\rho,A_\sigma-\partial_\sigma\lambda).
\label{applyGaugeTransf}
\end{eqn}
(We note that ${\partial_{A_\mu}}$ acts as a partial derivative, as expected. The space-time $X$ itself is invariant under such an `internal' gauge transformation, since $\v{v}_\lambda$---like $\v{v}_g$ in Eq.~(\ref{defnOfVg})---has no components of the form $\partial_{x^\mu}$. This is in contrast to a space-time translation $\partial_t$ or rotation ${y\partial_x-x\partial_y}$, for example.)

Given the EOM in Eq.~(\ref{vacMaxEOM}) and the symmetry characteristics in Eq.~(\ref{defGaugeSymm}), we now simply plug in for $\ws{E}_{u^\alpha}(\mL)$ and $Q^\alpha[\lambda]$ in Eq.~(\ref{N2TDiffIdentity}) to derive this system's N2T differential identity
\begin{eqn}
0&=\ws{E}_\lambda\Big[-(\partial_\sigma\lambda)\partial_\tau F^{\tau\sigma}\Big]\\
&=\partial_\sigma\partial_\tau F^{\tau\sigma}.
\label{exampleN2TIdentity}
\end{eqn}
As expected, because of the linearity of $\lambda(x)$ in $Q^{A_\mu}[\lambda]$, $\lambda(x)$ vanishes from Eq.~(\ref{exampleN2TIdentity}).

Eq.~(\ref{exampleN2TIdentity}) is the resultant N2T differential identity. Due to the antisymmetry of $F^{\mu\nu}$, this identity appears rather trivial, and conveys the appropriate sense that N2T produces off-shell identities independent of a system's dynamics. Nevertheless, merely from the gauge symmetry of $S$, the above N2T procedure sheds light on the kinematics of the Maxwell action. In the next section, we will find that the same procedure derives the local charge conservation law of the Vlasov-Maxwell system.

\section{Noether's second theorem for Vlasov-Maxwell systems\label{sectionN2T_Apply}}

\subsection{The continuous space-time Klimontovich-Maxwell model\label{sectionN2T_VM}}
We now use the preceding N2T procedure to systematically derive a charge conservation law for the continuous space-time Klimontovich-Maxwell system. This system specializes a Vlasov-Maxwell system to the following distribution function defined by $N$ point particles
\begin{eqn}
f(t,\v{x},\v{v})=\sum\limits_{j=1}^N\delta^{(3)}\big(\v{x}-\v{X}_j(t)\big)\delta^{(3)}\big(\v{v}-\dot{\v{X}}_j(t)\big).
\label{klimDist}
\end{eqn}
The Klimontovich-Maxwell system is accordingly described by the following action:
\begin{eqn}
S=\int\md^4x~\mL[\phi,\v{A},\v{X}_i]=\int\md^4x\Bigg[\frac{1}{2}&\big(\nabla\phi+\partial_t{\v{A}}\big)^2-\frac{1}{2}\big(\nabla\times\v{A}\big)^2\\
&+\sum_{j=1}^N\delta_j\cdot\left(\frac{1}{2}m_j\dot{\v{X}}_j^2-q_j\phi+q_j\v{A}\cdot\dot{\v{X}}_j\right)\Bigg].
\label{LForVMContinuous}
\end{eqn}
Here, ${\v{A}=\v{A}(t,\v{x})}$ is the vector potential, ${\phi=\phi(t,\v{x})}$ is the electric potential, ${\v{X}_i=\v{X}_i(t)}$ are particle positions and particle mass and charge are denoted by $m_i$ and $q_i$, respectively. We have also used the following shorthand for the delta function:
\begin{eqn}
\delta_j\defeq\delta^{(3)}\big(\v{x}-\v{X}_j(t)\big).
\end{eqn}

We apply Euler operators to derive the Euler-Lagrange equations of each field
\begin{eqn}
\ws{E}_\phi(\mL)&=\nabla\cdot\v{E}-\rho\\
\ws{E}_\v{A}(\mL)&=\partial_t\v{E}-\nabla\times\v{B}+\v{J}\\
\ws{E}_{\v{X}_i}(\mL)&=\delta_i\cdot\left[-m_i\ddot{\v{X}}_i+q_i\big(\v{E}+\dot{\v{X}}_i\times\v{B}\big)\right]
\label{EOMforKliMax}
\end{eqn}
where we have used the distributional derivative
\begin{eqn}
\int f(\eta)\delta'(\eta)\md\eta=-\int f'(\eta)\delta(\eta)\md\eta
\label{distrDeriv}
\end{eqn}
with $\eta\in\{t,\v{x}\}$, and where
\begin{eqn}
\v{E}(t,\v{x})&\defeq-\nabla\phi(t,\v{x})-\partial_t{\v{A}(t,\v{x})}\\
\v{B}(t,\v{x})&\defeq\nabla\times\v{A}(t,\v{x})\\
\rho(t,\v{x})&\defeq\sum_{j=1}^Nq_j\delta_j\\
\v{J}(t,\v{x})&\defeq\sum_{j=1}^Nq_j\dot{\v{X}}_j(t)\delta_j.
\label{fieldDefinitions}
\end{eqn}
As noted in Eq.~(\ref{DefineEInSitu}), an Euler operator $\ws{E}_u$ for an arbitrary field $u$ is essentially defined to accommodate integration by parts, such as that in Eq.~(\ref{distrDeriv}). In particular, total derivatives in $\mL$---e.g. $(f\delta)'$---that contribute to boundary terms of the action integral ${S=\int\mL\md^4x}$---e.g. $f\delta|_{-\infty}^\infty$---lie in the kernel of $\ws{E}_u$, such that ${\ws{E}_u(\mL+\ws{Div}\gamma)=\ws{E}_u(\mL)}$. Indeed, the operator relation ${\ws{E}_u\circ\ws{Div}=0}$ always holds \citep[see the `variational complex' of][]{olver_applications_1986}, where ${\ws{Div}}$ denotes a divergence.

We now note that the action of Eq.~(\ref{LForVMContinuous}) is invariant under the following gauge transformation:
\begin{eqn}
\phi\rightarrow&\phi'=\phi+\partial_t\lambda\\
\v{A}\rightarrow&\v{A}'=\v{A}-\nabla\lambda.
\label{contGaugeTransfs}
\end{eqn}
In particular, the electromagnetic terms of the Lagrangian are invariant, while the coupled particle terms pick up a divergence---namely, ${\mL\rightarrow\mL+\partial_\mu\gamma^\mu}$, where ${\gamma^\mu=-\sum_jq_j\delta_j\lambda\cdot(1,\dot{\v{X}}_j)}$---that vanishes on the boundary of $S$. The vector field corresponding to this transformation---equivalent to Eq.~(\ref{defGaugeSymm})---is given by
\begin{eqn}
\v{v}_\lambda=\sum\limits_\alpha Q^\alpha[\lambda]\partial_{u^\alpha}=(\partial_t\lambda)\partial_\phi-(\nabla\lambda)\cdot\partial_\v{A}
\label{vectorFieldGaugeContForm}
\end{eqn}
for an arbitrary smooth function $\lambda(x)$.

Finally, given EOM in Eq.~(\ref{EOMforKliMax}) and the characteristics of our gauge symmetry in Eq.~(\ref{vectorFieldGaugeContForm}), we may derive the differential identity of N2T using the construction of Eq.~(\ref{N2TDiffIdentity})
\begin{eqn}
\sum\limits_\alpha Q^\alpha[\lambda]\ws{E}_{u^\alpha}(\mL)&=Q^\phi[\lambda]\cdot\ws{E}_\phi(\mL)+Q^\v{A}[\lambda]\cdot\ws{E}_{\v{A}}(\mL)\\
&=(\partial_t\lambda)\big[\nabla\cdot\v{E}-\rho\big]-\nabla\lambda\cdot\big[\partial_t\v{E}-\nabla\times\v{B}+\v{J}\big]
\label{AssembleContinuousCaseN2T}
\end{eqn}
such that
\begin{eqn}
0&=\ws{E}_\lambda\bigg[\sum\limits_\alpha Q^\alpha[\lambda]\ws{E}_{u^\alpha}(\mL)\bigg]\\
&=-\partial_t\big[\nabla\cdot\v{E}-\rho\big]+\nabla\cdot\big[\partial_t\v{E}-\nabla\times\v{B}+\v{J}\big]\\
&=\partial_t\rho+\nabla\cdot\v{J}.
\label{consLawChargeKM}
\end{eqn}
In the final line, we have noted the equality of mixed partials and the vanishing divergence of the curl.

The N2T differential identity arising from the Klimontovich-Maxwell Lagrangian's local gauge symmetry evidently discovers the charge conservation law itself. By construction, this conservation law must hold off-shell and identically; in particular, Eq.~(\ref{consLawChargeKM}) does not require the equations of motion in order to hold true. It is a trivial conservation law---also referred to as a `strong' or `improper' conservation law \citep{brading_noethers_2000}---an often-overlooked fact that is immediately verified upon examining the definitions of $\rho$ and $\v{J}$ in Eq.~(\ref{fieldDefinitions}).

\subsection{The geometric PIC method of \citet{squire_geometric_2012}\label{sectionN2T_PIC}}

We now derive an analogous charge conservation law for the discrete, gauge-symmetric Vlasov-Maxwell PIC method defined by \citet{squire_geometric_2012}. In this PIC scheme, space-time is discretized by a $\text{$d$-dimensional}$ spatial simplicial complex (comprised of triangles in two dimensions or tetrahedra in three dimensions) whose structure is held constant throughout a uniformly discretized time. The time dimension may be envisaged as forming temporal edges that extend orthogonally from the spatial simplices, as in a triangular prism. We denote this ($d$+1)-dimensional prismal complex $P_C$. We use DEC \citep{desbrun_discrete_2005} to define fields on $P_C$ that are single-valued on its $k$-cells (or their circumcentric duals) for ${0\leq k\leq d+1}$. In the present paper, we shall assume a spatial dimensionality ${d=3}$, such that $P_C$ is four-dimensional, with three-dimensional spatial tetrahedra comprising each time slice.

We first review some elements of DEC formalism that are necessary in the present study. It will be useful to distinguish the spatial edges from the temporal edges of $P_C$, so we denote a vertex of $P_C$ by $\sm{i}{n}$, where $i$ is the spatial index of the vertex and $n$ is its temporal index. A discrete $0$-form $\alpha$ is then defined by its values at each vertex, and a discrete $1$-form $\beta$ by its values on each edge
\begin{eqn}
\alpha&=\sum_{\sm{i}{n}}\alpha^i_n\Delta^i_n\\
\beta&=\sum_{i,n}\beta^i_{n-\frac{1}{2}}\Delta^i_{n-\frac{1}{2}}+\sum_{[ij],n}\beta^{ij}_n\Delta^{ij}_n.
\label{cochain_bases}
\end{eqn}
Here we have expressed the discrete forms (equivalently, cochains) $\alpha$ and $\beta$ in terms of their cochain bases, where $\Delta^i_n$ is an element of the 0-cochain basis that maps $\sm{i}{n}$ to 1 and all other vertices to 0; $\Delta^{ij}_n$ is similarly an element of the 1-cochain basis that maps the oriented edge $[ij]$ to 1 and all others to 0. (A temporal edge is understood to be oriented in the positive time direction, and its cochain is denoted $\Delta^i_{n-\frac{1}{2}}$.) Discrete $k$-forms of higher degree may be constructed with cochain bases in essentially the same way. The formalism of cochain bases will prove especially useful when we derive EOM for the dynamical fields on $P_C$---that is, when we define a DEC Euler operator.

Let us denote the set of all vertices in $P_C$ by $\{v\}$, the set of spatial and temporal edges by ${\{e\}=\{e_s\}\sqcup\{e_t\}}$, and the set of spatial and `spatio-temporal' faces by ${\{f\}=\{f_s\}\sqcup\{f_t\}}$. The DEC exterior derivative $\md$, satisfying ${\md^2=0}$, may be defined \citep{elcott_building_2005,desbrun_discrete_2006} by a matrix multiplication in the cochain basis. For a 1-form $\beta$, we see this as follows:
\begin{eqn}
\md\beta=\md\left(\beta_e\Delta^e\right)=\beta_e\md\Delta^e=\beta_eW^e_f\Delta^f
\label{exteriorDerivByMatrix}
\end{eqn}
where the matrix entry $W^e_f$ stores the weight---$\{\pm1,0\}$---of the 1-cochain $\Delta^e$ in the 2-cochain $\Delta^f$. (We recall that the boundary operator on chains---$\partial$---is similarly determined by ${W^f_e={(W^e_f)}^T}$.) We adopt the Einstein summation convention in Eq.~(\ref{exteriorDerivByMatrix}) and hereafter for prismal complex indices: $\{v\}$, $\{e\}$, and $\{f\}$.

For example, the electromagnetic gauge field $A$---a discrete 1-form defined on all edges of $P_C$---neatly splits in into an electric potential ${\phi^i_{n-\frac{1}{2}}\defeq-A^i_{n-\frac{1}{2}}}$ and a vector potential $\v{A}^{ij}_n\defeq A^{ij}_n$, as follows:
\begin{eqn}
A&=-\sum_{i,n}\phi^i_{n-\frac{1}{2}}\Delta^i_{n-\frac{1}{2}}+\sum_{[ij],n}\v{A}^{ij}_n\Delta^{ij}_n\\
&=-\phi_{e_t}\Delta^{e_t}+\v{A}_{e_s}\Delta^{e_s}.
\label{defGaugeFieldAOnPC}
\end{eqn}
Using Eq.~(\ref{exteriorDerivByMatrix}), we may correspondingly express $\md A$ as
\begin{eqn}
\md A&=-\phi_{e_t}\md\Delta^{e_t}+\v{A}_{e_s}\md\Delta^{e_s}\\
&=\left(-\phi_{e_t}W^{e_t}_{f_t}+\v{A}_{e_s}W^{e_s}_{f_t}\right)\Delta^{f_t}+\v{A}_{e_s}W^{e_s}_{f_s}\Delta^{f_s}\\
&=E\wedge\md t+B\\
&=F,
\label{Adecomp}
\end{eqn}
where we have made use of the 1-form ${\md t\defeq\sum_{e_t}\Delta^{e_t}}$ and wedge product to implicitly define the spatial 1- and 2-forms $E$ and $B$, respectively, and the Faraday 2-form $F$ \citep{stern_geometric_2015}. (As a note of caution, we emphasize that the preceding vector potential $\v{A}$ is a single number on each spatial edge, and its bold notation is only suggestive. On the other hand, the Whitney interpolant of $\v{A}$ will coordinatize $\mR^3$ and thereby extend the single valued $\v{A}$ from the spatial edges of $P_C$ to a 3-component vector field, as we shall see.)

The map from primal $k$-forms to dual ${(4-k)}$-forms on $P_C$ is effected via the metric-dependent Hodge star operator, $\star$. The Hodge star is defined \citep{abraham_manifolds_1988,desbrun_discrete_2005} such that the symmetric, metric-induced inner product $(\cdot,\cdot)$ on $k$-forms satisfies
\begin{eqn}
(\omega,\nu)\mu=\omega\wedge{\star}\nu
\label{innerProdWedgeHodge}
\end{eqn}
for primal $k$-forms $\omega$ and $\nu$ and volume top form $\mu$. For a $k$-form $\omega$ on an $n$-dimensional cell complex (${n=4}$ on $P_C$), it can therefore be shown that
\begin{eqn}
\star(\star\omega)=(-1)^{k(n-k)+\text{Ind}(g)}\omega.
\label{starstar}
\end{eqn}
Here, ${\text{Ind}(g)=\#\{\ms{eig}[g]<0\}}$ represents the index of the metric $g$. In the present context, we adopt a $(-$$+$$+$$+)$ convention for our Lorentzian metric ${g=\eta}$, such that ${\text{Ind}(g)=1}$ on $P_C$. The dual ${(4-k)}$-form ${\star\alpha}$ is thus defined on a dual chain ${(\star\sigma)^{4-k}}$ as follows:
\begin{eqn}
\langle\star\alpha,\star\sigma\rangle=\epsilon(\sigma)\frac{\abs{\star\sigma}}{\abs{\sigma}}\langle\alpha,\sigma\rangle,
\label{defineStarEvaluation}
\end{eqn}
where
\begin{eqn}
\epsilon(\sigma)=
\begin{cases}
+1&\text{if $\sigma$ is entirely spacelike} \\
-1&\text{otherwise}.
\end{cases}
\end{eqn}
Here, ${\abs{\sigma^k}}$ denotes the $k$-volume of the $k$-dimensional $\sigma^k$ (where ${\abs{\sigma^0}=1}$ for a single vertex), and $\langle\cdot,\cdot\rangle$ denotes a $k$-cochain evaluated on a $k$-chain in Eq.~(\ref{defineStarEvaluation}).

Integration by parts on $P_C$---which is necessary for the derivation of EOM, as in Eq.~(\ref{DefineEInSitu})---may be facilitated via the codifferential operator, $\delta$. Up to boundary contributions, $\delta$ is a formal adjoint to $\md$, that is
\begin{eqn}
(\md\alpha,\beta)\mu=(\alpha,\delta\beta)\mu+\md(\alpha\wedge\star\beta).
\label{IBPforPrismalComplex}
\end{eqn}
When acting on a $k$-form defined on an $n$-dimensional complex, $\delta$ is given by \citep{abraham_manifolds_1988,desbrun_discrete_2005}
\begin{eqn}
\delta=(-1)^{n(k-1)+1+\text{Ind}(g)}\star\md\star.
\label{deltaDefn}
\end{eqn}
We observe that, whereas $\md$ maps a $k$-form to a $(k+1)$-form, $\delta$ reduces its degree to a $(k-1)$ form.

Having briefly reviewed relevant elements of DEC, we may now restate the discrete action of \citet{squire_geometric_2012}, defined on $P_C$
\begin{eqn}
S&=\sum_{V_{\sigma^2}}-\frac{1}{2}\md A\wedge\star\md A~+~\sum_{p,n}\left\{\frac{h}{2}m_p\left|\frac{\v{X}^p_{n+\frac{1}{2}}-\v{X}^p_{n-\frac{1}{2}}}{h}\right|^2~-~q_p\sum_{i}\phi^i_{n-\frac{1}{2}}\vphi^i\left(\v{X}^p_{n-\frac{1}{2}}\right)\right.\\
&\hspace{120pt}+~q_p\left(\frac{\v{X}^p_{n+\frac{1}{2}}-\v{X}^p_{n-\frac{1}{2}}}{h}\right)\cdot\sum_{[ij]}\v{A}^{ij}_n\int\limits_{t_{n-\frac{1}{2}}}^{t_{n+\frac{1}{2}}}\md t~\vphi^{ij}\left(\v{X}^p(t)\right)\Biggr\}.
\label{thePICAction}
\end{eqn}
In Eq.~(\ref{thePICAction}), we have denoted a sum over support volumes $V_{\sigma^2}$ for the primal-dual 4-form $\md A\wedge\star\md A$, with $A$ defined as in Eq.~(\ref{defGaugeFieldAOnPC}). $V_{\sigma^2}$ represents the convex hull of the 2-chain $\sigma^2$ and its dual $\star\sigma^2$ on which $\langle\md A,\sigma^2\rangle$ and $\langle\star\md A,\star\sigma^2\rangle$ are respectively defined. The symbol $h$ denotes the time step, $n$ the time index and $p$ the particle index. $\v{X}^p(t)$ is defined as the constant velocity path between the particle's staggered-time positions $\v{X}^p_{n-\frac{1}{2}}$ and $\v{X}^p_{n+\frac{1}{2}}$. In particular, particle paths are chosen to have straight line trajectories between the staggered times ${t\in\left[(n-\frac{1}{2})h,(n+\frac{1}{2})h\right]}$, $\forall$ ${n\in\mZ}$.

The Whitney 0-form $\vphi^i(x)$ and 1-form $\vphi^{ij}(x)$ interpolate $\phi$ and $\v{A}$ to an arbitrary point ${x\in P_C}$ \citep{bossavit_whitney_1988}. In effect, $\vphi^i$ and $\vphi^{ij}$ complete the spatial components of the cochain bases ${\Delta^i_n}$ and ${\Delta^{ij}_n}$ adopted in Eq.~(\ref{cochain_bases}) by extending DEC forms to the convex hull of $P_C$. In the continuous space-time of the Klimontovich-Maxwell system, the everywhere-defined gauge fields $\phi(t,\v{x})$ and $\v{A}(t,\v{x})$ were `attached' to point particles by the delta function, ${\delta^{(3)}(\v{x}-\v{X}_j(t))}$. In the prismal complex $P_C$, Whitney forms play this role by interpolating the gauge fields to the locations of point particles. Likewise, while we continue to avoid ascribing any geometric notion to point particles themselves, we see that Whitney forms on $P_C$ attach geometry to the charge densities and currents of the particles, as did the delta function in Eq.~(\ref{fieldDefinitions}). For example, the spatial dot product in Eq.~(\ref{thePICAction}) composes ${\v{X}^p_{n+\frac{1}{2}}}$ with the Whitney-interpolated 3-component vector field ${\v{A}^{ij}_n\vphi^{ij}(\v{X}^p(t))}$, (where $\v{A}^{ij}_n$ represents a single number and $\vphi^{ij}$ a 3-component vector).

We now follow the continuous space-time N2T procedure of Eqs.~(\ref{EOMforKliMax})-(\ref{consLawChargeKM}) by examining the equations of motion and gauge symmetry of $S$ in Eq.~(\ref{thePICAction}). As we have already seen, the differential structure of space-time has been replaced in the discrete setting by the prismal complex $P_C$, its DEC operators and Whitney forms. To derive the Euler-Lagrange equations of $S$, therefore, we must define an Euler operator for fields defined on this discrete structure. By analogy with Eq.~(\ref{EulerOperatorDef}), such an operator must implement a variational derivative on the space of fields defined on $P_C$.

Consider, for example, a $k$-form $\alpha$ defined by its expansion in $k$-cochain basis elements: ${\alpha=\sum_\sigma\alpha_\sigma\Delta^\sigma}$, where $\sigma$ ranges over all $k$-cells on $P_C$. Since each component $\alpha_\sigma$ of $\alpha$ can be varied independently, the variational derivative of $\alpha_\sigma$ takes the simple form of a partial derivative. We therefore define the Euler operator ${\ws{E}_{\alpha_\sigma}}$ on the action $S$ as follows:
\begin{eqn}
\ws{E}_{\alpha_\sigma}(S)\defeq&\pd{S}{\alpha_\sigma}\\
=&\pd{}{\alpha_\sigma}\sum L.
\label{EulerOperatorDefn}
\end{eqn}
(We note that in a discrete setting, variational derivatives are made to act on the entire action $S$, rather than on the Lagrangian, because discrete Lagrangians are necessarily non-local.) As usual we will assume that all fields and their variations have compact support, such that any divergence term in $L$---which contributes to ${S=\sum L}$ only at the boundary---vanishes under ${\ws{E}_{\alpha_\sigma}}$. Eq.~(\ref{EulerOperatorDefn}) is the DEC counterpart to the continuous Euler operator defined in Eq.~(\ref{EulerOperatorDef}), and is now applied to derive our EOM.

To calculate ${\ws{E}_{\phi_{e_t}}}(S)$ and ${\ws{E}_{\v{A}_{e_s}}}(S)$, we first re-express ${\md A\wedge\star\md A}$ in $S$ using Eqs.~(\ref{innerProdWedgeHodge})-(\ref{deltaDefn}) and the invariance of $S$ under the addition of a divergence (${L\rightarrow L+\md\gamma}$)
\begin{eqn}
\md A\wedge\star\md A\stackrel{(\ref{innerProdWedgeHodge})}{=}(\md A,\md A)\mu&\stackrel{(\ref{IBPforPrismalComplex})}{\approx}(A,\delta\md A)\mu\\
&\stackrel{(\ref{innerProdWedgeHodge})}{=}A\wedge\star\delta\md A\stackrel{(\ref{deltaDefn})}{=}A\wedge\star\star\md\star\md A\stackrel{(\ref{starstar})}{=}A\wedge\md\star\md A,
\end{eqn}
where $\approx$ indicates equality up to an (ignorable) divergence. Then, using Eq.~(\ref{defGaugeFieldAOnPC}) and noting the symmetry of the intermediate expression ${(\md A,\md A)\mu}$ above, we apply the Euler operator of Eq.~(\ref{EulerOperatorDefn}) to derive the EOM for $A$ as follows:
\begin{subequations}
\begin{alignat}{3}
0&=\ws{E}_{\phi_{e_t}}(S)~&=&&\Delta^{e_t}\wedge\md\star\md A-\rho^{e_t}
\label{eqOfMotionPIC1}\\
0&=\ws{E}_{\v{A}_{e_s}}(S)~&=&-&\Delta^{e_s}\wedge\md\star\md A+J^{e_s}
\label{eqOfMotionPIC2}
\end{alignat}
\end{subequations}
where
\begin{subequations}
\begin{align}
\rho^{e_t}&\defeq\sum_pq_p\vphi^i\left(\v{X}^p_{t(e_t)}\right)\label{defnChargeDEC}\\
J^{e_s}&\defeq\sum_pq_p\left(\frac{\v{X}^p_{t_f(e_s)}-\v{X}^p_{t_i(e_s)}}{h}\right)\cdot\int\limits_{t_i(e_s)}^{t_f(e_s)}\md t~\vphi^{e_s}\left(\v{X}^p(t)\right).\label{defnCurrentDEC}
\end{align}
\end{subequations}
In Eq.~(\ref{defnChargeDEC}), $i$ denotes the spatial vertex associated with $e_t$, and $\v{X}^p_{t(e_t)}$ denotes the position of particle $p$ at the time coincident with the midpoint $\sm{i}{n-\frac{1}{2}}$ of $e_t$. In Eq.~(\ref{defnCurrentDEC}), $\v{X}^p_{t_i(e_s)}$ and $\v{X}^p_{t_f(e_s)}$ denote the initial and final particle positions, respectively, coincident with the midpoints $\sm{i}{n-\frac{1}{2}}$ and $\sm{i}{n+\frac{1}{2}}$ that bookend the ${t=n}$ time slice containing $e_s$.

It is worth pausing to interpret these EOM. We first observe that the primal-dual wedge product in Eq.~(\ref{eqOfMotionPIC1}) is only non-vanishing on the spatial ${(\star\Delta^{e_t})}$ component of ${\md\star\md A}$. This follows from the definition of the primal-dual wedge product, which is only non-zero on the convex hulls of a cell and its dual: ${CH(\sigma,\star\sigma)}$. Reading off from Eq.~(\ref{Adecomp}), therefore, Eq.~(\ref{eqOfMotionPIC1}) becomes
\begin{eqn}
\rho^{e_t}=\Delta^{e_t}\wedge\md\star\left(E\wedge\md t\right)=\md D\wedge\Delta^{e_t},
\end{eqn}
Gauss's law for the electric displacement dual 2-form $D$, as expected. An analogous interpretation of Eq.~(\ref{eqOfMotionPIC2}) yields a discrete Amp\`ere-Maxwell law. We have omitted the ${\ws{E}_{\v{X}^p}(S)}$ EOM for particle trajectories, as they will not be necessary for the derivation of charge conservation via N2T---just as they were unnecessary in Eqs.~(\ref{AssembleContinuousCaseN2T})-(\ref{consLawChargeKM}). These implicit time-step particle EOM are derived in \citet{squire_geometric_2012}.

%The geometric interpretation of Whitney forms is a subject of recent research (e.g. \citep{salamon_geometric_2014}), but we simply observe that in our treatment above the Whitney `1-form' is more appropriately construed as a primal-dual 4-form. As a final comment, w

Having derived our field equations of motion, we must now examine the gauge symmetry of the action $S$ in Eq.~(\ref{thePICAction}). In particular, $S$ is invariant under the local gauge transformation ${A\rightarrow A-\md f}$, defined by
\begin{eqn}
\phi^i_{n+\frac{1}{2}}&~\rightarrow~\phi^i_{n+\frac{1}{2}}+\delta\phi^i_{n+\frac{1}{2}}=\phi^i_{n+\frac{1}{2}}+\left(f^i_{n+1}-f^i_n\right)\\
\v{A}^{ij}_n&~\rightarrow~\v{A}^{ij}_n+\delta\v{A}^{ij}_n=\v{A}^{ij}_n-\left(f^j_n-f^i_n\right),
\label{discGaugeTransfs}
\end{eqn}
where ${f=f_v\Delta^v}$ is an arbitrary primal 0-form on $P_C$.

After all, the electromagnetic term of $S$---${\md A\wedge\star\md A}$---is clearly invariant under ${A\rightarrow A-\md f}$, since ${\md^2=0}$. Furthermore, as noted in \citet{squire_geometric_2012}, the gauge invariance of the particle terms of $S$ follows from a defining property of Whitney interpolation: ${\md_{\text{c}}((\alpha)_\text{interp})=(\md_{\text{d}}\alpha)_{\text{interp}}}$, where ${\md_{\text{c}}}$ and ${\md_{\text{d}}}$ denote continuous and discrete exterior derivatives, respectively and ${(\cdot)_{\text{interp}}}$ denotes Whitney interpolation. Eq.~(\ref{discGaugeTransfs}) therefore transforms $L\rightarrow L+\md_c\gamma$, adding a divergence term analogous to the transformation of Eq.~(\ref{contGaugeTransfs}) for the continuous space-time Vlasov-Maxwell system.

Following Eq.~(\ref{vectorFieldGaugeContForm}), the gauge transformation of Eq.~(\ref{discGaugeTransfs}) is seen to be generated by a vector field
\begin{eqn}
\v{v}_f&=\sum\limits_\alpha Q^\alpha[f]\partial_{u^\alpha}=\sum\limits_{e_t}\left(\md_{e_t}f\right)\partial_{\phi_{e_t}}-\sum\limits_{e_s}\left(\md_{e_s}f\right)\partial_{\v{A}_{e_s}},
\label{defDiscGaugeSymm}
\end{eqn}
where the coefficient ${\md_ef=f_{v_2}-f_{v_1}}$ corresponds to the oriented edge ${e=[v_1v_2]}$, and where the sums are taken over all temporal and spatial edges, respectively.

We have thus gathered the necessary data to complete the N2T construction of Eq.~(\ref{N2TDiffIdentity}) for our DEC system. As in Eq.~(\ref{AssembleContinuousCaseN2T}), we note
\begin{eqn}
\sum\limits_\alpha Q^\alpha[f]\ws{E}_{u^\alpha}(S)&=\sum\limits_{e_t}Q^{\phi_{e_t}}[f]\cdot\ws{E}_{\phi_{e_t}}(S)+\sum\limits_{e_s}Q^{\v{A}_{e_s}}[f]\cdot\ws{E}_{\v{A}_{e_s}}(S)\\
&=\sum\limits_{e_t}\left(\md_{e_t}f\right)\cdot\Big(\Delta^{e_t}\wedge\md\star\md A-\rho^{e_t}\Big)\\
&\hspace{10pt}+\sum\limits_{e_s}\left(-\md_{e_s}f\right)\cdot\Big(-\Delta^{e_s}\wedge\md\star\md A+J^{e_s}\Big).
\label{setupConsLaw}
\end{eqn}
We now observe that
\begin{eqn}
(\md_ef)\Delta^e=\md(f_v\Delta^v)=\md f,
\end{eqn}
so applying the Euler operator for $f_v$ at vertex ${v={\sm{i}{n}}}$ to Eq.~(\ref{setupConsLaw}) yields
\begin{eqn}
0&=\ws{E}_{f_v}\left[\sum\limits_\alpha Q^\alpha[f]\ws{E}_{u^\alpha}(S)\right]\\
&=\ws{E}_{f_v}\left[\sum\limits_{V_{\sigma^1}}\md f\wedge\md\star\md A-\sum\limits_{e_t}\left(\md_{e_t}f\right)\cdot\rho^{e_t}-\sum\limits_{e_s}\left(\md_{e_s}f\right)\cdot J^{e_s}\right]\\
&=\Delta^v\wedge\star\delta\star\md\star\md A+\left(\rho^i_{n+\frac{1}{2}}-\rho^i_{n-\frac{1}{2}}\right)+\sum\limits_jJ^{[ij]}\\
&=\left(\rho^i_{n+\frac{1}{2}}-\rho^i_{n-\frac{1}{2}}\right)+\sum\limits_jJ^{[ij]}
\label{SimplicialConsLaw}
\end{eqn}
since up to a sign, ${(\delta\star\md)=(\star\hspace{1pt}\md\hspace{1pt}{\star\hspace{1pt}\star}\hspace{1pt}\md)={\star}\hspace{1pt}\md^2=0}$. The sum of $J^{[ij]}$ over $j$ captures all spatial edges that terminate on vertex ${v={\sm{i}{n}}}$.

The last equality of Eq.~(\ref{SimplicialConsLaw}) reveals the desired charge conservation law on $P_C$. By its very construction through N2T, this conservation law is guaranteed to be an off-shell differential identity, as was Eq.~(\ref{consLawChargeKM}). We readily verify this fact as follows.

First, we restrict our sources $\rho$ and $J$ to a particle of charge $q$ whose path over one time step remains within a single spatial tetrahedron; the general case follows without significant alteration. We then recall \citep{bossavit_whitney_1988} that the Whitney 0-form $\vphi^i$ interpolates from vertex $i$ via barycentric coordinates such that, over the tetrahedron $[ijk\ell]$,
\begin{eqn}
\vphi^i+\vphi^j+\vphi^k+\vphi^\ell=1.
\label{Whitney0Identity}
\end{eqn}
In vector form, the 1-form $\vphi^{ij}$ is then given by
\begin{eqn}
\vphi^{ij}=\vphi^i\nabla\vphi^j-\vphi^j\nabla\vphi^i.
\end{eqn}
Summing over the three spatial edges terminating on vertex $i$ of the tetrahedron containing the particle, therefore
\begin{eqn}
\sum\limits_{j\neq i}\vphi^{ij}&=\vphi^i\nabla\left(\sum\limits_{j\neq i}\vphi^j\right)-\left(\sum\limits_{j\neq i}\vphi^j\right)\nabla\vphi^i\\
&=\vphi^i\nabla\left(1-\vphi^i\right)-\left(1-\vphi^i\right)\nabla\vphi^i\\
&=-\nabla\vphi^i.
\end{eqn}

It follows, then, that
\begin{eqn}
\sum\limits_{j\neq i}J^{[ij]}&=q\left(\frac{\v{X}_f-\v{X}_i}{h}\right)\cdot\int\limits_{t_i}^{t_f}\md t\sum\limits_{j\neq i}\vphi^{ij}\left(\v{X}(t)\right)\\
&=-q\left(\frac{\v{X}_f-\v{X}_i}{h}\right)\cdot\int\limits_{t_i}^{t_f}\md t\nabla\vphi^i\left(\v{X}(t)\right)\\
&=-q\int_i^f\v{v}\intprod\md\vphi^i\\
&=-q\left[\vphi^i(\v{X}_f)-\vphi^i(\v{X}_i)\right]
\end{eqn}
where $\v{v}\intprod\md\vphi^i$ is the interior product of the exact form $\md\vphi^i$ with respect to the velocity ${\v{v}\defeq\frac{1}{h}(\v{X}_f-\v{X}_i)}$, which is constant over a single time step of the particle. Upon comparison with the definition for $\rho$ in Eq.~(\ref{defnChargeDEC}), it is clear that Eq.~(\ref{SimplicialConsLaw}) holds off-shell, as desired. The N2T formalism of \citet{hydon_extensions_2011} has succeeded in deriving the off-shell, discrete conservation law.

Before we depart from the Lagrangian formalism, we note that an alternative approach to deriving the conservation laws of the continuous space-time and DEC Vlasov-Maxwell actions---Eqs.~(\ref{LForVMContinuous}) and (\ref{thePICAction})---entails gauge fixing these actions by setting ${\phi(x)=0}$ and ${\phi^i_{n+\frac{1}{2}}=0}$, respectively. Such a gauge fixing removes these systems' degeneracy and uniquely determines solutions to their equations of motion. In such an approach, N1T is applied to the time-independent symmetry transformation ${\v{A}(t,\v{x})\rightarrow\v{A}(t,\v{x})-\nabla\psi(\v{x})}$, thereby deriving a non-trivial conservation law in the form of a time evolution of Gauss's law. In the ensuing sections, we pursue an analogous gauge-fixing approach for the Hamiltonian Vlasov-Maxwell system, employing the Hamiltonian formalism's counterpart to N1T---the momentum map.

\section{The momentum map and reduction of the Vlasov-Maxwell system\label{sectionSympRed}}

Having derived the N2T charge conservation laws of continuous and discrete Vlasov-Maxwell Lagrangian systems, we now explore the conservation laws of these gauge-symmetric systems in the Hamiltonian formalism. We first develop the necessary technical background for Sections~${\ref{sectionMomMap}\text{-}\ref{sectionCanonCons_PIC}}$, which study gauge-compatible splitting methods and their application to PIC algorithms. In this section, we review the Poisson structure of the Vlasov-Maxwell system, derived in \citet{morrison_maxwell-vlasov_1980} and independently in \citet{iwinski_canonical_1976}, and later presented in its complete, correct form in \citet{marsden_hamiltonian_1982}. Closely following this last reference, we review the Poisson reduction \citep{marsden_reduction_1974,marsden_reduction_1986} of the Vlasov-Maxwell system, which `spends' the system's gauge symmetries in order to eliminate their associated redundant (gauge) degrees of freedom. As we will discuss at length, this Poisson reduction is achieved via the momentum map, which in turn determines the local charge conservation law of the Vlasov-Maxwell system. The following section serves as a concise pedagogical summary of \citet{marsden_hamiltonian_1982}, with additional discussion relevant to the more recent plasma physics literature.

\subsection{The Poisson structure of the Vlasov-Maxwell system}

We first recall the Poisson bracket of \citet{marsden_hamiltonian_1982} for the Vlasov-Maxwell system,
\begin{eqn}
\{\{F,G\}\}[f,\v{A},\v{Y}]=&\int\md\v{x}\md\v{p}~f\left\{\frac{\delta F}{\delta f},\frac{\delta G}{\delta f}\right\}_{\v{x}\v{p}}+\int\md\v{x}\left(\frac{\delta F}{\delta\v{A}}\cdot\frac{\delta G}{\delta\v{Y}}-\frac{\delta G}{\delta\v{A}}\cdot\frac{\delta F}{\delta\v{Y}}\right)
\label{UnredPoissonBracket}
\end{eqn}
with time evolution defined by the Hamiltonian
\begin{eqn}
H[f,\v{A},\v{Y}]=&\frac{1}{2}\int f\cdot\abs{\v{p}-\v{A}}^2\md\v{x}\md\v{p}+\frac{1}{2}\int\Big[\abs{\v{Y}}^2+\abs{\nabla\times\v{A}}^2\Big]\md\v{x}.
\label{UnredHamilDef}
\end{eqn}
Here, $F$ and $G$ represent arbitrary functionals of the distribution function ${f(\v{x},\v{p})}$, the 3-component vector potential ${\v{A}(\v{x})}$ and its conjugate momentum ${\v{Y}(\v{x})}$. As we shall see momentarily, $\v{Y}$ can be readily identified as negative the electric field strength---(i.e., ${\v{Y}=-\v{E}}$). We note that our system is rendered in the temporal gauge, wherein the electric potential satisfies ${\phi(\v{x})=0}$. As in \cite{marsden_hamiltonian_1982}, we denote the Poisson bracket in Eq.~(\ref{UnredPoissonBracket}) by $\{\{\cdot,\cdot\}\}$ merely to distinguish it from other Poisson structures.

The ${\int\md\v{x}\md\v{p}~f\{\delta_f\cdot,\delta_f\cdot\}_{\v{xp}}}$ operator in the first line of Eq.~(\ref{UnredPoissonBracket}) is a Lie-Poisson bracket \citep{marsden_introduction_1999}, which defines a Poisson structure for functions on a dual Lie algebra $\mgd$. In general, the Lie-Poisson bracket on an arbitrary dual Lie algebra $\mgd$ is defined to inherit the bracket ${[\cdot,\cdot]}$ of its underlying Lie algebra $\mg$ as follows:
\begin{eqn}
\{F,G\}(\alpha)\defeq-\left\langle\alpha,\left[\frac{\delta F}{\delta \alpha},\frac{\delta G}{\delta \alpha}\right]\right\rangle.
\label{LiePoiss}
\end{eqn}
The bracket of Eq.~(\ref{LiePoiss}) is defined $\forall$ ${F,G\in C^\infty(\mgd)}$ with respect to some fixed ${\alpha\in\mgd}$, where ${\langle\cdot,\cdot\rangle}$ represents the linear pairing of elements of $\mgd$ and $\mg$. The functional derivative ${{\delta F/\delta \alpha}\in\mgdd}$ can be seen to define a linear function on $\mgd$, in that it acts as a directional derivative on the functional $F$ at the `point' $\alpha\in\mgd$. In particular, for arbitrary $\beta\in\mgd$
\begin{eqn}
\left\langle\beta,\frac{\delta F}{\delta \alpha}\right\rangle=D_\alpha F\cdot\beta=\left.\d{}{\epsilon}\right|_{\epsilon=0}F(\alpha+\epsilon\beta).
\end{eqn}
Since ${\mgdd\cong\mg}$, the functional derivative may be interpreted as an element of the Lie algebra.

In the present context, the Lie algebra $\mg$ corresponds to infinitesimal transformations of ${(\v{x},\v{p})\cong\mR^6}$, the position-momentum phase space. Such transformations can be regarded as Hamiltonian vector fields on $\mR^6$, which map via anti-homomorphism to their corresponding generating functions, i.e. 
\begin{eqn}
[\v{X}_h,\v{X}_k]=-\{h,k\}_{\v{xp}}.
\label{antihomoMorph}
\end{eqn}
The bracket ${\{\cdot,\cdot\}_{\v{xp}}}$ therefore serves as a Lie bracket, defined pointwise on $\mR^6$
\begin{eqn}
\{h,k\}_{\v{xp}}\defeq\Big(\partial_\v{x}h\cdot\partial_\v{p}k-\partial_\v{x}k\cdot\partial_\v{p}h\Big).
\label{lieBracketFcnGen}
\end{eqn}
The dual Lie algebra $\mgd$ is similarly identified by distribution densities on $\mR^6$, which pair linearly to Hamiltonian functions via integration
\begin{eqn}
\big\langle f,h\big\rangle\defeq\int fh~\md\v{x}\md\v{p}
\label{linPairR6}
\end{eqn}
for ${f\in\mgd,h\in\mg}$.

In this way, the operator ${\int\md\v{x}\md\v{p}~f\{\delta_f\cdot,\delta_f\cdot\}_{\v{xp}}}$ comprising the first term of Eq.~(\ref{UnredPoissonBracket}) is seen to be a Lie-Poisson bracket of the form in Eq.~(\ref{LiePoiss}). We note that the negative sign of Eq.~(\ref{LiePoiss}) cancels with the negative sign of the anti-homomorphism of Eq.~(\ref{antihomoMorph}) to produce this operator.

The second term of Eq.~(\ref{UnredPoissonBracket}) represents the electromagnetic `sector' of our Poisson structure, and derives from the canonical symplectic structure on the cotangent space---${T^*Q=\{(\v{A},\v{Y})\}}$---of the configuration space ${Q=\{\v{A}\}}$. Therefore, the complete setting of the Vlasov-Maxwell system is a Poisson manifold, given by
\begin{eqn}
M=\mgd\times T^*Q
\end{eqn}
with its bracket defined in Eq.~(\ref{UnredPoissonBracket}).

We now consider dynamics on this Poisson manifold $M$. To derive our Hamiltonian EOM, it is convenient to define functionals
\begin{eqn}
F(\v{u})\defeq\int\md\v{u}'F(\v{u}')\delta(\v{u}-\v{u}')
\label{defineFunctional}
\end{eqn}
for $F\in\{f,\v{A},\v{Y}\}$ as in \citet{kraus_gempic:_2017}, where $\v{u}=(\v{x},\v{p})$ or $\v{u}=\v{x}$, as appropriate. Plugging such functionals into Eqs.~(\ref{UnredPoissonBracket})-(\ref{UnredHamilDef}), we find
\begin{eqn}
\dot{f}(\v{x},\v{p})=\{\{f,H\}\}&=-\Big[\partial_\v{x}f+\partial_\v{p}f\cdot(\nabla\v{A})\Big]\cdot(\v{p}-\v{A})\\
\dot{\v{A}}(\v{x})=\{\{\v{A},H\}\}&=\v{Y}\\
\dot{\v{Y}}(\v{x})=\{\{\v{Y},H\}\}&=\int f\cdot(\v{p}-\v{A})\md\v{p}-\nabla\times\nabla\times\v{A}.
\label{unredEOM}
\end{eqn}

We observe that $\v{Y}$ plays the role of $-\v{E}$, as expected. For convenience, we note that the familiar form of the Vlasov equation may be recovered from the first line of Eq.~(\ref{unredEOM}) by defining a distribution density $\bar{f}$ on ${(\v{x},\v{v})}$ space where ${\v{v}=\v{p}-\v{A}}$, i.e.
\begin{eqn}
f(\v{x},\v{p})=\bar{f}(\v{x},\v{p}-\v{A})=\bar{f}(\v{x},\v{v}),
\end{eqn}
such that ${\partial_\v{x}f=\partial_\v{x}\bar{f}-(\nabla\v{A})\cdot\partial_\v{v}\bar{f}}$; ${\partial_\v{p}f=\partial_\v{v}\bar{f}}$; and ${\dot{f}=\partial_t\bar{f}-\dot{\v{A}}\cdot\partial_\v{v}\bar{f}}$. Here, we use ${\nabla\equiv\partial_\v{x}}$ interchangeably, and adopt the dyad convention
\begin{eqn}
\v{v}\cdot\v{A}\v{B}\cdot\v{w}=v_iA_iB_jw_j
\end{eqn}
in Einstein notation.

\subsection{Gauge symmetry and the momentum map}

With our Poisson and Hamiltonian structure in hand, we now examine the gauge symmetry of the Vlasov-Maxwell system. Continuing to follow \citet{marsden_hamiltonian_1982}, we define a group action $\Phi_\psi:M\rightarrow M$ on our Poisson manifold ${M=\mgd\times T^*Q}$ of the form
\begin{eqn}
\Phi_\psi:(f,\v{A},\v{Y})\mapsto\Big(f\circ\tau_\psi,~\v{A}-\nabla\psi,~\v{Y}\Big),
\label{defGroupAction}
\end{eqn}
where
\begin{eqn}
\tau_\psi(\v{x},\v{p})\defeq(\v{x},\v{p}+\nabla\psi).
\label{tauDef}
\end{eqn}
We emphasize that $\Phi_\psi$ transforms $f$, and not $\v{p}$ itself. It is straightforward to check that $\Phi_\psi$ is a \emph{canonical group action}, i.e. that the Poisson bracket is preserved by the pullback of $\Phi_\psi$, namely ${\Phi_\psi^*\{\{F,G\}\}=\{\{\Phi_\psi^*F,\Phi_\psi^*G\}\}}$.

We define such an arbitrary function ${\psi\in\mF}$ as belonging to the abelian group ${\mF\defeq C^\infty(\mR^3)}$ of smooth functions on $\mR^3$, with the group composition law of addition. Its Lie algebra $\mf$ is also identifiable as the smooth functions on $\mR^3$, while its dual $\mfd$ is the set of densities over $\mR^3$ that pair to elements of $\mf$ via integration over $\mR^3$---analogous to the $\mR^6$ integration of Eq.~(\ref{linPairR6}).

Now let ${\phi\in\mf}$ denote an arbitrary Lie algebra element, such that ${\exp(\epsilon\phi)\in\mF}$ $\forall$ ${\epsilon\in\mR}$. By differentiating the group action ${\Phi_{\exp(\epsilon \phi)}}$ on $M$, we may associate to any such ${\phi\in\mf}$ the corresponding vector field ${\phi_M}$ on $M$, namely
\begin{eqn}
\phi_M\defeq\left.\d{}{\epsilon}\right|_{\epsilon=0}\Phi_{\exp(\epsilon \phi)}.
\label{definePhiM}
\end{eqn}
The vector field $\phi_M$ is therefore the infinitesimal generator of the group action on $M$ corresponding to ${\phi\in\mf}$.

For any canonical group action on Poisson manifold $M$, we may seek a corresponding \emph{momentum map}.
The momentum map ${\mu:M\rightarrow\mfd}$ of a group action is defined such that, $\forall$ $\phi\in\mf$ and ${m\in M}$, the induced function
\begin{eqn}
\mu^\phi:M&\rightarrow\mR\\
m&\mapsto\langle \mu(m),\phi\rangle
\end{eqn}
satisfies
\begin{eqn}
\{\{F,\mu^\phi\}\}=\phi_M(F)
\label{defMomMap}
\end{eqn}
for arbitrary ${F\in C^\infty(M)}$. Here, $\phi_M(F)$ is the Lie derivative of $F$ along the vector field $\phi_M$. In particular, the momentum map $\mu$ assigns a dual element of $\mfd$ to each point of $M$ such that, when $\mu$ is everywhere paired with an element ${\phi\in\mf}$ of the Lie algebra, the resulting function $\mu^\phi$ on $M$ is a generating function of the associated vector field $\phi_M$. 

The preceding definition of $\mu$ is general to arbitrary Poisson systems with canonical group actions, and we now apply it to find $\mu$ for the Vlasov-Maxwell system of interest. We first note that a single point ${m\in M=\mgd\times T^*Q}$ specifies $(f,\v{A},\v{Y})$ over the entire $(\v{x},\v{p})$ phase space. Given the group action defined in Eqs.~(\ref{defGroupAction})-(\ref{tauDef}), it is immediately seen that $\phi_M$ can be expressed as the following infinitesimal operator on $M$ corresponding to $\phi(\v{x})\in\mf$:
\begin{eqn}
\{\{\cdot,\mu^\phi\}\}=&\int\md\v{x}\md\v{p}~\nabla\phi\cdot\pd{f}{\v{p}}\frac{\delta}{\delta f}-\int\md\v{x}~\nabla\phi\cdot\frac{\delta}{\delta\v{A}}.
\label{JphiOperator}
\end{eqn}
Upon inspection, it is evident that to generate the operator of Eq.~(\ref{JphiOperator}), the Poisson bracket of Eq.~(\ref{UnredPoissonBracket}) requires that $\mu^\phi$ be given by
\begin{eqn}
\mu^\phi(m)\defeq&\langle \mu(m),\phi\rangle\\
=&\int\md\v{x}\left[\int\md\v{p}~f(\v{x},\v{p})+\nabla\cdot\v{Y}\right]\phi(\v{x}),
\label{muPhiDef}
\end{eqn}
where ${\langle\cdot,\cdot\rangle=\int\md\v{x}}$. Therefore, the momentum map must be
\begin{eqn}
\mu(m)=&\int\md\v{p}~f(\v{x},\v{p})+\nabla\cdot\v{Y}\\
\defeq&\rho+\nabla\cdot\v{Y},
\label{unredMomMap}
\end{eqn}
where ${\rho(\v{x})\defeq\int\md\v{p}f(\v{x},\v{p})}$. We note that ${\mu^\phi:M\rightarrow\mR}$ is a real-valued function while ${\mu(m)\in\mfd}$ is a density on $\mR^3$, as desired.

For later use, we further observe that $\mu$ is group equivariant:
\begin{eqn}
\mu\circ\Phi_\psi=\text{Ad}_{\psi^{-1}}^*\circ\mu,
\label{muEquivariance}
\end{eqn}
where $\text{Ad}_{\psi^{-1}}^*$ represents the coadjoint action \citep{marsden_introduction_1999} of $\psi\in\mF$ on an element of $\mfd$. In particular, it is clear by inspection of Eq.~(\ref{unredMomMap}) that $\mu$ is invariant under $\mF$ transformations, and since $\mF$ is abelian, its coadjoint action on $\mfd$ is simply the identity map.

\subsection{Deriving the conservation law}

The momentum map $\mu$ is defined as above---Eqs.~(\ref{definePhiM})-(\ref{defMomMap})---for any Poisson manifold $M$ with a canonical group action $\Phi$. If it should happen that a Hamiltonian $H$ is furthermore defined on $M$ such that $H$ is invariant under $\Phi$, then the momentum map $\mu$ so-constructed further guarantees a conservation law for the system.

Let us show this for our Vlasov-Maxwell system. We first note that $\Phi_\psi$ leaves the Hamiltonian invariant, ${\Phi_\psi^*H=H}$. By differentiating this expression with respect to $\psi$ as in Eq.~(\ref{definePhiM}), it is seen that, infinitesimally
\begin{eqn}
0=\phi_M(H)=\{\{H,\mu^\phi\}\}=-\{\{\mu^\phi,H\}\}=-\dot{\mu}^\phi.
\label{infinitesimalMuInv}
\end{eqn}
Each linearly independent ${\phi\in\mf}$ therefore determines a unique first integral of the system---i.e., $\mu^\phi$.

We can make a stronger observation as well. Since ${\dot{\mu}^\phi=0}$ holds for arbitrary ${\phi\in\mf}$, the entire momentum map is invariant under the flow of $H$, that is
\begin{eqn}
\dot{\mu}=\{\{{\mu,H\}}\}=0.
\label{muConservation}
\end{eqn}
This follows rigorously from the fundamental lemma of variational calculus applied to $\dot{\mu}^\phi$ via Eq.~(\ref{muPhiDef}). As a result, we can apply the definition of Eq.~(\ref{unredMomMap}) to derive
\begin{eqn}
0=\dot{\mu}&=\dot{\rho}+\nabla\cdot\dot{\v{Y}}.
\label{hamilConsLaw}
\end{eqn}
This completes the canonical derivation of the Vlasov-Maxwell local conservation law---${\dot{\mu}=0}$---in the continuous Hamiltonian formalism. We note that, setting ${\v{Y}=-\v{E}}$, Eq.~(\ref{hamilConsLaw}) is the time evolution of Gauss's law.

With an additional substitution to Eq.~(\ref{hamilConsLaw}) from the EOM for $\dot{\v{Y}}$ in Eq.~(\ref{unredEOM}), we may re-express this canonical conservation law in the form
\begin{eqn}
0=\dot{\rho}+\nabla\cdot\v{J},
\label{usualHamilChargeConsLaw}
\end{eqn}
where ${\v{J}\defeq\int f\cdot(\v{p}-\v{A})\md\v{p}}$. Here $\rho$ and $\v{J}$ are (scalar and vector) densities over $\mR^3$ and functionals in the sense of Eq.~(\ref{defineFunctional}). This charge conservation law may be immediately checked by substituting the expression for $\dot{f}$ from Eq.~(\ref{unredEOM}). In the present Hamiltonian context, it is evident that Eq.~(\ref{usualHamilChargeConsLaw}) can no longer be regarded as an off-shell identity. (After all, time evolution itself is only `dynamically defined', so to speak, by the Hamiltonian.)

\subsection{Reduction of the Vlasov-Maxwell system\label{subsectionReduceVM}}

Finally, we undertake the Poisson reduction \citep{marsden_reduction_1974,marsden_reduction_1986} of the Vlasov-Maxwell system. Given a Poisson manifold ${(M,~\{\cdot,\cdot\}_M)}$ on which a Lie group $G$ acts by Poisson diffeomorphisms, the Poisson reduction of $M$ is the unique quotient map ${\pi:(M,~\{\cdot,\cdot\}_M)\rightarrow (M/G,~\{\cdot,\cdot\}_{M/G})}$ satisfying ${\pi^*\{f,g\}_{M/G}=\{\pi^*f,\pi^*g\}_{M}}$. For a Poisson system equipped with a group-equivariant momentum map $\mu$ satisfying Eq.~(\ref{muEquivariance})---as in the Vlasov-Maxwell system of interest---such a quotient map may be defined on level sets of $\mu$, as we now describe.

Consider the preimage of an arbitrary ${\alpha\in\mfd}$ under ${\mu:M\rightarrow\mfd}$, that is, the level set ${\mu^{-1}(\alpha)\subset M}$. We may take equivalence classes of this preimage under the full action of $\Phi_\psi$ $\forall$ $\psi\in\mF$. That is, we reduce ${\mu^{-1}(\alpha)}$ to the quotient manifold ${\mu^{-1}(\alpha)/\mF}$, and thereby take a `slice' of the orbit of ${\mu^{-1}(\alpha)}$ under the action of $\mF$. Because $\mu$ is equivariant in the sense of Eq.~(\ref{muEquivariance}), this quotient is well defined. The reduced manifold ${\mu^{-1}(\alpha)/\mF}$ will again be a Poisson manifold, as we now show.

Let us consider the particular case ${\alpha=0}$, and define ${M_0\defeq\mu^{-1}(0)}$. By Eq.~(\ref{unredMomMap}), $M_0$ corresponds to the submanifold of $M$ on which ${\rho=-\nabla\cdot\v{Y}}$. We now take equivalence classes of $M_0$ under the orbit of $\mF$ by defining new phase space coordinates that are invariant under the action of Eq.~(\ref{defGroupAction}), namely
\begin{eqn}
\bar{f}(\v{x},\v{v})&=f(\v{x},\v{p}=\v{v}+\v{A})\\
\v{B}&=\nabla\times\v{A}\\
\v{E}&=-\v{Y}.
\label{newM0Coor}
\end{eqn}
We therefore identify the manifold of equivalence classes ${\tilde{M}_0\defeq M_0/\mF}$ with the manifold $(\bar{f},\v{B},\v{E})$ of densities $\bar{f}$ defined on ${(\v{x},\v{v})}$ space, vector fields $\v{B}$ that satisfy ${\nabla\cdot\v{B}=0}$, and vector fields $\v{E}$ that satisfy ${\bar{\rho}=\nabla\cdot\v{E}}$, where now ${\bar{\rho}\defeq\int\bar{f}~\md\v{v}}$. (We note that the choice to constrain M to ${\{m\in M~|~\mu(m)=0\}}$ evidently corresponds to the physical case ${\nabla\cdot\v{E}-\bar{\rho}=0}$, in which no `external' charges are present in the system. Such a choice must be made when determining the Vlasov-Maxwell system's initial conditions, as we shall see.) Our reduction map is therefore summarized by
\begin{eqn}
\pi_{\text{red}}:~~~~~~~~~~~~~~~~\mu^{-1}(0)\subset M~~~~&\longrightarrow~~~\tilde{M}_0\defeq\mu^{-1}(0)/\mF\\
\big(f(\v{x},\v{p}),\v{A},\v{Y}\big)~~~~&\longmapsto~~~\big(\bar{f}(\v{x},\v{v}),\v{B},\v{E}\big).
\end{eqn}

As calculated in \citet{marsden_hamiltonian_1982} Sec.~7, the substitution of Eq.~(\ref{newM0Coor}) into the bracket of Eq.~(\ref{UnredPoissonBracket}) yields the following reduced Poisson bracket on $\tilde{M}_0$:
\begin{eqn}
\{\{F,G\}\}^{\text{red}}[\bar{f},\v{B},\v{E}]=&\int\md\v{x}\md\v{v}~\Bigg[\bar{f}\left\{\frac{\delta F}{\delta\bar{f}},\frac{\delta G}{\delta\bar{f}}\right\}_{\v{x}\v{v}}+\\
&\hspace{-95pt}\bar{f}\v{B}\cdot\left(\pd{}{\v{v}}\frac{\delta F}{\delta\bar{f}}\times\pd{}{\v{v}}\frac{\delta G}{\delta\bar{f}}\right)+\left(\frac{\delta F}{\delta\v{E}}\cdot\pd{\bar{f}}{\v{v}}\frac{\delta G}{\delta\bar{f}}-\frac{\delta G}{\delta\v{E}}\cdot\pd{\bar{f}}{\v{v}}\frac{\delta F}{\delta\bar{f}}\right)\Bigg]\\
&\hspace{-80pt}+\int\md\v{x}\left(\frac{\delta F}{\delta\v{E}}\cdot\nabla\times\frac{\delta G}{\delta\v{B}}-\frac{\delta G}{\delta\v{E}}\cdot\nabla\times\frac{\delta F}{\delta\v{B}}\right).
\label{RedPoissonBracket}
\end{eqn}

We note that this process of Poisson reduction preserves the ${\dot{\mu}=0}$ conservation law associated with the unreduced Poisson manifold $M$. After all, the image of $\pi_{\text{red}}$ restricts $M$ to (quotients of) a submanifold ${M_0\subset M}$ on which $\mu$ is constant---in particular, level sets of a single value of $\mu$. The conservation law of Eq.~(\ref{hamilConsLaw}) is clearly respected by this reduction, and may simply be re-expressed in the phase space variables of the reduced manifold $\tilde{M}_0$, along with its form in Eq.~(\ref{usualHamilChargeConsLaw}), i.e.
\begin{eqn}
0=\dot{\bar{\mu}}&=\dot{\bar{\rho}}-\nabla\cdot\dot{\v{E}}\\
&=\dot{\bar{\rho}}+\nabla\cdot\bar{\v{J}},
\end{eqn}
where
\begin{eqn}
\bar{\mu}&=\int\md\v{v}~\bar{f}(\v{x},\v{v})-\nabla\cdot\v{E}\\
&=\bar{\rho}-\nabla\cdot\v{E},
\end{eqn}
and where ${\bar{\rho}\defeq\int\bar{f}~\md\v{v}}$ and ${\bar{\v{J}}\defeq\int\bar{f}\v{v}~\md\v{v}}$.

We note that the reduced bracket of Eq.~(\ref{RedPoissonBracket}) is a well-defined Poisson bracket specifically on the quotient submanifold $\tilde{M}_0$. Some of the plasma physics literature \citep[e.g.][]{morrison_poisson_1982,morrison_general_2013,kraus_gempic:_2017} notes that Eq.~(\ref{RedPoissonBracket}) generally fails to satisfy a Jacobi identity, however, so we pause to elucidate the source of this contrasting point of view.

In particular, the aforementioned literature defines the Vlasov-Maxwell system on an augmented manifold that includes all unconstrained vector fields ${\v{E},\v{B}\in\mR^3}$:
\begin{eqn}
\tilde{M}_0^+\defeq\tilde{M}_0\sqcup\left\{\v{E},\v{B}~|~\nabla\cdot\v{E}\neq\bar{\rho},\nabla\cdot\v{B}\neq0\right\}.
\label{AugmentViaDisjointUnion}
\end{eqn}
When the bracket of Eq.~(\ref{RedPoissonBracket}) is defined on $\tilde{M}_0^+$ and not on $\tilde{M}_0$, it no longer everywhere obeys the Jacobi identity \citep{morrison_poisson_1982,chandre_use_2013}; in particular, the Jacobi identity is satisfied on $\tilde{M}_0^+$ only when ${\nabla\cdot\v{B}=0}$. Indeed, the constraint ${\nabla\cdot\v{B}=0}$ appears as an exogenous defect that must be satisfied for ${(\tilde{M}_0^+,~\{\{\cdot,\cdot\}\}^{\text{red}})}$ to be considered a Poisson manifold. On $\tilde{M}_0^+$, the bracket of Eq.~(\ref{RedPoissonBracket}) also acquires additional Casimirs,
\begin{eqn}
\{\{\cdot,\bar{\rho}-\nabla\cdot{\v{E}}\}\}&=0\\
\{\{\cdot,\nabla\cdot\v{B}\}\}&=0,
\label{casimirFormulas}
\end{eqn}
in much the same way that a Poisson structure on ${\mR^2=\{(x,y)\}}$ acquires a $z$ Casimir when the system is embedded in $\mR^3$.

We adopt the point of view that it is more natural to regard the bracket of Eq.~(\ref{RedPoissonBracket}) as a Poisson bracket defined on the submanifold of physical interest---${\tilde{M}_0=\mu^{-1}(0)/\mF}$---rather than a defected bracket defined on the larger manifold including arbitrary vector fields $\v{E}$ and $\v{B}$. In a sense, it is merely a lack of economical notation that leads us to coordinatize $\tilde{M}_0$ with vector symbols $\v{E}$ and $\v{B}$ that are more commonly defined over all of $\mR^3$.

It is clear from this discussion, however, that care must be taken in any numerical implementation of the reduced Vlasov-Maxwell bracket to constrain one's fields to the phase space of $\tilde{M}_0$; generic, unconstrained vector fields ${\v{E},\v{B}\in\mR^3}$ are to be avoided.

\section{The momentum map in Hamiltonian splitting methods\label{sectionMomMap}}

We now reconsider the momentum map---and its associated conservation laws---in the context of Hamiltonian splitting algorithms. Due to their ease of computation, splitting methods offer an appealing algorithmic implementation of many Hamiltonian systems \citep[for example, see ][]{he_hamiltonian_2016}. In effect, a splitting method splits a system's Hamiltonian $H$ into some finite number of `sub-Hamiltonians' ${\{H_i\}}$ such that
\begin{eqn}
H=\sum\limits_{i=1}^NH_i.
\end{eqn}
The system's dynamical variables $\v{u}$ are then evolved by each subsystem individually, arranged in a sequence chosen to minimize discretization error, e.g.
\begin{eqn}
\hspace{-5pt}\v{u}(t+\Delta t)&=\exp\Big(\Delta tH\Big)\v{u}(t)\\
&\approx\exp\left(\frac{\Delta t}{2}H_1\right)\exp\Big(\Delta tH_2\Big)\exp\left(\frac{\Delta t}{2}H_1\right)\v{u}(t),
\end{eqn}
where we have schematically represented two subsystems, ${H=H_1+H_2}$, arranged in a second-order Strang splitting \citep{hairer_geometric_2006}.

The advantage afforded by this subdivision of the Hamiltonian is that its subsystems $\{H_i\}$ are often more easily integrated individually than the full system $H$. In fact, each sub-Hamiltonian $H_i$ can at times be made sufficiently simple to allow its exact integration, without any discrete approximation. We will see examples of this exact evolution in the Vlasov-Maxwell splitting methods detailed in Section~\ref{sectionCanonCons_PIC}.

Our interest concerns the status of the momentum map $\mu$ in such algorithms. A sufficient condition for the exact preservation of $\mu$ in splitting methods is, in fact, quite straightforward to state. In particular, let us suppose that each sub-Hamiltonian is gauge invariant---that is, invariant under the group action of some group $G$
\begin{eqn}
\Phi_g^*H_i=H_i~~~~~~~~\forall~i\text{ and }g\in G.
\label{HiGroupInvariance}
\end{eqn}
Then, differentiating with respect to $g$---as in Eq.~(\ref{definePhiM})---we find by the same argument of Eqs.~(\ref{infinitesimalMuInv})-(\ref{muConservation}) that ${\dot{\mu}=0}$ in each Hamiltonian subsystem, where $\mu$ is the total system's momentum map associated with the group action $\Phi_g$.

This claim follows simply from the observation that $\mu$ is an object defined by its Poisson manifold ${(M,\{\cdot,\cdot\})}$, separate and apart from the Hamiltonian defined on that manifold. (One might say that $\mu$ is defined kinematically \citep{morrison_hamiltonian_1993} on $M$, independently of dynamics.) $\mu$ is therefore preserved along the flow generated by any gauge-invariant function. Consequently, if each sub-Hamiltonian is gauge invariant and its flow is exactly integrated, then the momentum map is exactly preserved by its evolution during each discrete time step. We summarize this result as a theorem.

\textbf{Theorem.} Let $\Phi$ be a canonical group action of a Lie group $G$ on Poisson manifold $M$ with momentum map $\mu$, and let ${H:M\rightarrow\mR}$ satisfy ${\Phi_g^* H=H}$, $\forall$ $g\in G$. Suppose a splitting method ${H=\sum_{i=1}^NH_i}$ satisfies:
\begin{enumerate}[leftmargin=.75cm,labelsep=0cm,align=left,label={(\arabic*)}]\vspace{4pt}
\item~~~${\Phi_g^* H_i=H_i}$, $\forall$ $i$ and $g\in G$;\vspace{4pt}
\item~~~subsystem $H_i$ is solved exactly $\forall$ $i$.\vspace{4pt}
\end{enumerate}
Then $\mu$ is exactly preserved by the splitting method---that is, ${\dot{\mu}=0}$.

We refer to such an algorithm as a \emph{gauge-compatible splitting method}. Gauge-compatible splitting methods have a distinct advantage over other time discretizations of Hamiltonian systems, in that they preserve the geometric structure of the systems they simulate (in particular, the momentum map) and therefore obey exact conservation laws.

\section{Conservation laws in Hamiltonian PIC splitting methods\label{sectionCanonCons_PIC}}

\subsection{An `unreduced' Hamiltonian PIC method\label{sectionCanonCons_PIC1}}

With the formalism we have developed, we proceed to explore PIC methods in the Hamiltonian setting, by defining a PIC splitting method adapted from \citet{xiao_explicit_2015} and \citet{qin_canonical_2016}. The latter of these references implements a symplectic-Euler integrator for the unreduced Poisson bracket of Eq.~(\ref{UnredPoissonBracket}), while the former implements a splitting method for the reduced bracket of Eq.~(\ref{RedPoissonBracket}). We shall synthesize the two, defining a gauge-compatible splitting method for the unreduced bracket of Eq.~(\ref{UnredPoissonBracket}) and, in so doing, demonstrate the merit of this new class of splitting methods. The result is an explicit-time-advance, canonical, locally charge-conserving PIC method, whose momentum map and conservation law we shall systematically derive.

In \citet{qin_canonical_2016}, a Klimontovich-Maxwell PIC method is derived from the unreduced bracket of Eq.~(\ref{UnredPoissonBracket}) by specifying the following form for the distribution function $f(\v{x},\v{p})$ of $L$ particles, analogous to Eq.~(\ref{klimDist})
\begin{eqn}
f(\v{x},\v{p})=\sum\limits_{i=1}^L\delta^{(3)}(\v{x}-\v{X}_i)\delta^{(3)}(\v{p}-\v{P}_i).
\label{discrete_f}
\end{eqn}
Here, ${(\v{X}_i,\v{P}_i)}$ denotes the dynamical coordinates of particle $i$ in phase space. The fields $\v{A}$ and $\v{Y}$ are also discretized on a (three-dimensional) spatial lattice and are denoted ${(\v{A}_n,\v{Y}_n)}$ at lattice site $n$. We shall further require the interpolation of $\v{A}$
\begin{eqn}
\v{A}(\v{x})=\sum\limits_{n=1}^N\v{A}_nW_{\sigma_1}(\v{x}-\v{x}_n)
\label{discrete_A}
\end{eqn}
where $n$ is an index over all $N$ lattice sites and $W_{\sigma_1}$ is an (as yet unspecified) interpolation function for $\v{A}$.

A Poisson bracket for this discrete system simply follows from the canonical symplectic structure of its variables. In particular, we define the symplectic manifold
\begin{eqn}
M_d=T^*X\times T^*Q,
\end{eqn}
where ${X=\mR^{3L}}$ is the space of particle position coordinates and ${Q=\mR^{3N}}$ is the space of vector potentials on the lattice, such that ${T^*X=\{(\v{X}_i,\v{P}_i)\}}$ and ${T^*Q=\{(\v{A}_n,\v{Y}_n)\}}$. A point ${m\in M_d}$ correspondingly specifies ${(\v{X}_i,\v{P}_i,\v{A}_n,\v{Y}_n)}$ $\forall$ $i,n$ (where the subscript $d$ denotes discretization).

The Poisson bracket for this symplectic manifold therefore takes its usual Darboux-coordinate form
\begin{eqn}
\hspace{-10pt}\{\{F,G\}\}_d[\v{X}_i,\v{P}_i,\v{A}_n,\v{Y}_n]=&\sum\limits_{i=1}^L\left(\pd{F}{\v{X}_i}\cdot\pd{G}{\v{P}_i}-\pd{G}{\v{X}_i}\cdot\pd{F}{\v{P}_i}\right)\\
&+\sum\limits_{n=1}^N\left(\pd{F}{\v{A}_n}\cdot\pd{G}{\v{Y}_n}-\pd{G}{\v{A}_n}\cdot\pd{F}{\v{Y}_n}\right).
\label{NewBracket}
\end{eqn}
We observe that, unlike its continuous counterpart in Eq.~(\ref{UnredPoissonBracket}), the bracket of Eq.~(\ref{NewBracket}) is non-degenerate; it defines $M_d$ not only as a Poisson manifold, but as a symplectic manifold.

The discrete Hamiltonian of \citet{qin_canonical_2016} is derived from Eq.~(\ref{UnredHamilDef}) by substituting the Klimontovich distribution of Eq.~(\ref{discrete_f}) and expanding terms of the form ${\abs{\v{P}_i-\v{A}(\v{X}_i)}^2}$ using Eq.~(\ref{discrete_A})
\begin{eqn}
\hspace{-10pt}H_d[\v{X}_i,\v{P}_i,\v{A}_n,\v{Y}_n]&=\frac{1}{2}\sum\limits_{i=1}^L\Bigg[\v{P}_i^2-2\v{P}_i\cdot\sum\limits_{n=1}^N\v{A}_nW_{\sigma_1}(\v{X}_i-\v{x}_n)\\
&\hspace{-40pt}+\sum\limits_{m,n=1}^N\v{A}_m\cdot\v{A}_nW_{\sigma_1}(\v{X}_i-\v{x}_m)W_{\sigma_1}(\v{X}_i-\v{x}_n)\Bigg]+\frac{1}{2}\sum\limits_{n=1}^N\Big[\v{Y}_n^2+\left|\nabla_d^+\times\v{A}\right|_n^2\Big].
\label{Hamil_d}
\end{eqn}
Here the operator ${\left(\nabla_d^\pm\times\right)_n}$ represents a discrete curl, defined by
\begin{eqn}
\left(\nabla_d^\pm\times\v{A}\right)_n\defeq\pm\left(
\begin{matrix}
\frac{A^3_{i,j\pm1,k}-A^3_{i,j,k}}{\Delta y}-\frac{A^2_{i,j,k\pm1}-A^2_{i,j,k}}{\Delta z}\vspace{5pt}\\
\frac{A^1_{i,j,k\pm1}-A^1_{i,j,k}}{\Delta z}-\frac{A^3_{i\pm1,j,k}-A^3_{i,j,k}}{\Delta x}\vspace{5pt}\\
\frac{A^2_{i\pm1,j,k}-A^2_{i,j,k}}{\Delta x}-\frac{A^1_{i,j\pm1,k}-A^1_{i,j,k}}{\Delta y}\\
\end{matrix}
\right)
\end{eqn}
for ${n=(i,j,k)}$.

We now describe the gauge symmetry of this discrete Hamiltonian system. We define the group action $\Phi_f$ on $M_d$ by analogy with Eq.~(\ref{defGroupAction})
\begin{eqn}
\Phi_f(\v{X}_i,\v{P}_i,&\v{A}_n,\v{Y}_n)=\left(\v{X}_i,\Big[\v{P}_i-\nabla_d^+f(\v{X}_i)\Big],\Big[\v{A}_n-(\nabla_d^+f)_n\Big],\v{Y}_n\right),
\label{groupAction_d}
\end{eqn}
where
\begin{eqn}
\nabla_d^+f(\v{x})=\sum\limits_{n=1}^N(\nabla_d^+f)_nW_{\sigma_1}(\v{x}-\v{x}_n)
\end{eqn}
and where ${(\nabla_d^\pm)_n}$ is a discrete gradient defined by
\begin{eqn}
(\nabla_d^\pm f)_n\defeq\pm\left(
\begin{matrix}
\frac{f_{i\pm1,j,k}-f_{i,j,k}}{\Delta x}\vspace{5pt}\\
\frac{f_{i,j\pm1,k}-f_{i,j,k}}{\Delta y}\vspace{5pt}\\
\frac{f_{i,j,k\pm1}-f_{i,j,k}}{\Delta z}
\end{matrix}
\right).
\end{eqn}
We note that ${\nabla_d^\pm\times\nabla_d^\pm=0}$ as an operator. (If the $\pm$ signs agree, this relation holds identically; if they disagree, it holds only after a summation over lattice points, $\sum_n$.) We also note that---in contrast with Eqs.~(\ref{defGroupAction})-(\ref{tauDef})---$\v{P}_i$ and $\v{A}_n$ are shifted in the same direction in Eq.~(\ref{groupAction_d}), reflecting the fact that $\v{p}$ and $\v{P}_i$ have opposite signs in Eq.~(\ref{discrete_f}) when we reinterpret the transformation of Eq.~(\ref{tauDef}) as a transformation of $\v{P}_i$.

The function $f$ appearing in the group action of Eq.~(\ref{groupAction_d}) is to be understood as a scalar function defined only at lattice points. In particular, $f\in\mF_d$ is a group element of the set $\mF_d$ of discrete scalar functions with an abelian composition law of addition. Its Lie algebra $\mf_d$ is also the set of discrete scalar functions on the lattice, while its dual $\mfd_d$ is the set of densities, which pair to elements of $\mf_d$ by summing over pointwise products
\begin{eqn}
\big\langle\alpha,\phi\big\rangle\defeq\sum\limits_{n=1}^N\alpha_n\phi_n~~~\forall~\alpha\in\mfd_d,~\phi\in\mf_d.
\label{linPairR6_d}
\end{eqn}

We must verify that the group action is canonical, a task most easily approached infinitesimally. In particular, we ask whether the following infinitesimal form of ${\{\{\Phi_f^*F,\Phi_f^*G\}\}_d=\Phi_f^*\{\{F,G\}\}_d}$ holds:
\begin{eqn}
&\left\{\left\{-\nabla_d^+\phi(\v{X}_i)\cdot\pd{F}{\v{P}_i}-\nabla_d^+\phi_n\cdot\pd{F}{\v{A}_n}~,~G\right\}\right\}_d-(F\leftrightarrow G)\\
&\hspace{90pt}=-\nabla_d^+\phi(\v{X}_i)\cdot\pd{\{\{F,G\}\}_d}{\v{P}_i}-\nabla_d^+\phi_n\cdot\pd{\{\{F,G\}\}_d}{\v{A}_n},
\label{infinitesimalCondition}
\end{eqn}
where summation over repeated indices is implicit. After applying Eq.~(\ref{NewBracket}) to evaluate each bracket, Eq.~(\ref{infinitesimalCondition}) is seen to be true only when ${\nabla\times\nabla_d^+\phi(\v{X}_i)=0}$. This requires the operator relation
\begin{eqn}
\nabla\times\nabla_d^+=0.
\label{delOperatorConstraint}
\end{eqn}
Here, $\nabla\equiv\partial_{\v{X}_i}$ is a continuous spatial gradient.

Eq.~(\ref{delOperatorConstraint}) therefore necessitates the following condition on the interpolation function $W_{\sigma_1}$:
\begin{eqn}
\sum\limits_{n=1}^N(\nabla_d^+\phi)_nW_{\sigma_1}(\v{x}-\v{x}_n)=\nabla\sum\limits_{n=1}^N\phi_nW_{\sigma_0}(\v{x}-\v{x}_n)
\label{condOnW}
\end{eqn}
for some interpolation function $W_{\sigma_0}$. This condition was already discovered in \citet{xiao_explicit_2015}, and is analogous to a property of the simplicial Whitney forms described earlier \citep{bossavit_whitney_1988}. Our discussion of this condition merely contributes that, in a Hamiltonian context, the motivation for the constraint in Eq. (6.13) is the canonicality of the group action.

We now solve for $\mu_d$, the momentum map on $M_d$ associated with the group action of Eq.~(\ref{groupAction_d}), using the symplectic structure of Eq.~(\ref{NewBracket}). First, we must find the infinitesimal generator $\phi_{M_d}$ of our group action on $M_d$, defined analogously to Eq.~(\ref{definePhiM}). Given the group action of Eq.~(\ref{groupAction_d}) we expect $\phi_{M_d}$ to take the form (already implicitly used in Eq.~(\ref{infinitesimalCondition}))
\begin{eqn}
\{\{\cdot,\mu^\phi_d\}\}_d=&-\sum\limits_{i=1}^L\nabla_d^+\phi(\v{X}_i)\cdot\pd{}{\v{P}_i}-\sum\limits_{n=1}^N(\nabla_d^+\phi)_n\cdot\pd{}{\v{A}_n},
\label{JphiOperator_d}
\end{eqn}
where we denote the pairing of the momentum map with $\phi$ by ${\big\langle\mu_d,\phi\big\rangle=\mu^\phi_d}$. The Poisson bracket of Eq.~(\ref{NewBracket}) therefore requires that $\mu^\phi_d$ be given by
\begin{eqn}
\mu^\phi_d(m)&=\sum\limits_{n=1}^N(\nabla_d^+\phi)_n\cdot\left[\sum\limits_{i=1}^L\int_{-\infty}^{\v{X}_i}\md\v{X}_i'W_{\sigma_1}(\v{X}_i'-\v{x}_n)-\v{Y}_n\right]\\
&=\sum\limits_{n=1}^N\phi_n\nabla_d^-\cdot\left[-\sum\limits_{i=1}^L\int_{-\infty}^{\v{X}_i}\md\v{X}_i'W_{\sigma_1}(\v{X}_i'-\v{x}_n)+\v{Y}_n\right]
\label{unredPairedMomMap_d}
\end{eqn}
where in the second line we have summed by parts \citep{hydon_extensions_2011} using the discrete divergence operator
\begin{eqn}
\nabla_d^\pm\cdot\v{v}_n\defeq\pm\sum\limits_{\alpha=1}^3\frac{v^\alpha_{n\pm \hat{\alpha}}-v^\alpha_n}{\Delta x^\alpha}.
\end{eqn}
Note that $\md\v{X}_i'$ is treated in Eq.~(\ref{unredPairedMomMap_d}) and hereafter as a vector, with each component integrated individually. We observe that ${\nabla_d^\pm\cdot\nabla_d^\pm\times=0}$ as an operator (when $\pm$ signs agree).

Given the pairing defined in Eq.~(\ref{linPairR6_d}), the momentum map $\mu_d$ must therefore be
\begin{eqn}
\big(\mu_d(m)\big)_n=-\nabla_d^-\cdot\sum\limits_{i=1}^L\int_{-\infty}^{\v{X}_i}\md\v{X}_i'W_{\sigma_1}(\v{X}_i'-\v{x}_n)+\nabla_d^-\cdot\v{Y}_n
\label{mu_d}
\end{eqn}
defined at each lattice site ${n\in[1,N]}$. Due to the gauge invariance of $H_d$ in Eq.~(\ref{Hamil_d})---that is, ${\Phi_f^*H_d=H_d}$---the full system evolved in \emph{continuous} time by $H_d$ obeys the conservation law
\begin{eqn}
\dot{\mu}_d=0,
\label{discMuDotZero}
\end{eqn}
as in the continuous Vlasov-Maxwell system of Section~\ref{sectionSympRed}. Eqs.~(\ref{mu_d})-(\ref{discMuDotZero}) define the conservation law of our discrete Hamiltonian system in continuous time, systematically derived via the momentum map.

Following the analysis of Eqs.~(\ref{hamilConsLaw})-(\ref{usualHamilChargeConsLaw}), we may re-express this conservation law by deriving the continuous-time EOM of the full Hamiltonian $H_d$, as follows:
\begin{eqn}
\dot{\v{X}}_i&=\{\{\v{X}_i,H_d\}\}_d=\v{P}_i-\sum\limits_{m=1}^N\v{A}_mW_{\sigma_1}(\v{X}_i-\v{x}_m)\\
\dot{\v{P}}_i&=\{\{\v{P}_i,H_d\}\}_d=\sum\limits_{m=1}^N\Big(\dot{\v{X}}_i\cdot\v{A}_m\Big)\nabla W_{\sigma_1}(\v{X}_i-\v{x}_m)\\
\dot{\v{A}}_n&=\{\{\v{A}_n,H_d\}\}_d=\v{Y}_n\\
\dot{\v{Y}}_n&=\{\{\v{Y}_n,H_d\}\}_d=\sum\limits_{i=1}^L\dot{\v{X}}_iW_{\sigma_1}(\v{X}_i-\v{x}_n)-\left(\nabla_d^-\times\nabla_d^+\times\v{A}\right)_n.
\end{eqn}
Now substituting $\dot{\v{Y}}_n$ into the charge conservation law Eqs.~(\ref{mu_d})-(\ref{discMuDotZero}), we note that
\begin{eqn}
0&=\dot{\rho}_n+\nabla_d^-\cdot\v{J}_n
\label{splitLocChargeCons}
\end{eqn}
where
\begin{eqn}
\rho_n&\defeq-\nabla_d^-\cdot\sum\limits_{i=1}^L\int_{-\infty}^{\v{X}_i}\md\v{X}_i'W_{\sigma_1}(\v{X}_i'-\v{x}_n)\\
\v{J}_n&\defeq\sum\limits_{i=1}^L\dot{\v{X}}_iW_{\sigma_1}(\v{X}_i-\v{x}_n).
\label{chargeDefs}
\end{eqn}
This is an alternative form of the charge conservation law Eq.~(\ref{discMuDotZero}) for the continuous-time evolution of the Hamiltonian system of \citet{qin_canonical_2016}. (We observe that, unlike its counterpart in Eq.~(\ref{usualHamilChargeConsLaw}), it is an off-shell identity.)

The form of $\rho_n$ in Eq.~(\ref{chargeDefs}) can be justified by a schematic one-dimensional example in which ${W_{\sigma_1}(x)=1}$ on ${0\leq x<\Delta x}$ and 0 otherwise. For a single particle at ${X_i=0.2}$, we have
\begin{eqn}
\rho_n&=-\nabla_d^-\cdot\int_{-\infty}^{0.2}\md X_i'W_{\sigma_1}(X_i'-x_n)=
\begin{cases}
0.8/\Delta x&~n=0\\
0.2/\Delta x&~n=1\\
0&~n\neq0,1.
\end{cases}
\end{eqn}
This result demonstrates the appropriateness of the momentum map's systematically derived charge density.

We now define an algorithmic solution of this Hamiltonian system via a splitting method, and examine the preservation of $\mu_d$. To algorithmically evolve this system in discrete time, we define a gauge-compatible splitting method adapted from \citet{he_hamiltonian_2015,he_hamiltonian_2016}. We define Hamiltonian subsystems
\begin{eqn}
H_d=\sum\limits_{\alpha=1}^3H_{\text{Klim}}^\alpha+H_\v{A}+H_\v{Y}
\end{eqn}
where
\begin{eqn}
H_{\text{Klim}}^\alpha&\defeq\frac{1}{2}\sum\limits_{i=1}^L\Big(P_i^\alpha-A^\alpha(\v{X}_i)\Big)^2\\
H_\v{A}&\defeq\frac{1}{2}\sum\limits_{n=1}^N\left|\nabla_d^+\times\v{A}\right|_n^2\\
H_\v{Y}&\defeq\frac{1}{2}\sum\limits_{n=1}^N\v{Y}_n^2.
\end{eqn}
We immediately note that these subsystems are all gauge invariant for the group action of Eq.~(\ref{groupAction_d})---$\Phi_f^*H_i=H_i$ $\forall$ ${f\in\mF_d}$ and $i$---and will therefore comprise a gauge-compatible splitting---and preserve $\mu_d$---if they can be exactly solved.

Let us examine the EOM for each subsystem $H_i$ in turn
\begin{eqn}
H_{\text{Klim}}^\alpha
\begin{cases}
\dot{X}_i^\beta&=\delta_\alpha^\beta\left[P_i^\alpha-\sum\limits_{m=1}^NA_m^\alpha W_{\sigma_1}(\v{X}_i-\v{x}_m)\right]\\
\dot{P}_i^\beta&=\dot{X}_i^\alpha\sum\limits_{m=1}^NA_m^\alpha\partial_\beta W_{\sigma_1}(\v{X}_i-\v{x}_m)\\
\dot{A}_n^\beta&=0\\
\dot{Y}_n^\beta&=\delta^\beta_\alpha\sum\limits_{i=1}^L\dot{X}_i^\alpha W_{\sigma_1}(\v{X}_i-\v{x}_n)
\end{cases}
\label{KlimEOM}
\end{eqn}
\begin{eqn}
H_\v{A}
\begin{cases}
\dot{\v{X}}_i&=0\\
\dot{\v{P}}_i&=0\\
\dot{\v{A}}_n&=0\\
\dot{\v{Y}}_n&=-\left(\nabla_d^-\times\nabla_d^+\times\v{A}\right)_n
\end{cases}
\end{eqn}
\begin{eqn}
H_\v{Y}
\begin{cases}
\dot{\v{X}}_i&=0\\
\dot{\v{P}}_i&=0\\
\dot{\v{A}}_n&=\v{Y}_n\\
\dot{\v{Y}}_n&=0
\end{cases}
\end{eqn}
where ${\partial_\beta\equiv\partial/\partial X_i^\beta}$. (We emphasize that $\alpha$ is fixed, and is not summed over in the expressions for $H_{\text{Klim}}^\alpha$.) $H_\v{A}$ and $H_\v{Y}$ are exactly solvable at a glance. Furthermore, $H_{\text{Klim}}^\alpha$ is seen to be exactly solvable by noting that ${\ddot{X}_i^\beta=0}$; $\dot{X}_i^\beta$ is therefore a constant determined by a time step's initial conditions. The evolutions of $\dot{\v{P}}_i$ and $\dot{\v{Y}}_n$ in $H_{\text{Klim}}^\alpha$ follow immediately from this analysis.

The exact time evolutions of $H_\v{A}$, $H_\v{Y}$ and $H_{\text{Klim}}^\alpha$ are therefore explicitly solved, defining by construction an explicit-time-advance gauge-compatible splitting method that exactly preserves the momentum map, ${\dot{\mu}_d=0}$, as desired. The alternative form of the charge conservation law given in Eq.~(\ref{splitLocChargeCons})---that is, ${\dot{\rho}_n+\nabla_d^-\cdot\v{J}_n=0}$---is also exactly preserved in this algorithm, because the substitution that led from Eq.~(\ref{discMuDotZero}) to Eq.~(\ref{splitLocChargeCons})---that is, ${\nabla_d^-\cdot\dot{\v{Y}}_n=\nabla_d^-\cdot\v{J}_n}$---holds for each Hamiltonian subsystem above.

Finally, we note that the momentum map $\mu_d$ has significant ramifications for the appropriate initial conditions of the preceding algorithm. We refer the reader to a brief but important discussion of these initial conditions in the text following Eq.~(\ref{tildeMd0Constraints}) below.

\subsection{A `reduced' Hamiltonian PIC method\label{subSecRedHamPIC}}

We now examine the PIC method of \citet{xiao_explicit_2015}, which employs a splitting method equivalent to that of the preceding section for the reduced Vlasov-Maxwell bracket of Eq.~(\ref{RedPoissonBracket}).

We will mirror \citet{xiao_explicit_2015} and derive this PIC scheme by undertaking the symplectic reduction \citep{marsden_reduction_1974} of the discrete canonical bracket defined in Eq.~(\ref{NewBracket}). As in Section~\ref{subsectionReduceVM}, we define a mapping to the reduced symplectic manifold ${{\tilde{M}_d}_0=\mu_d^{-1}(0)/\mF_d}$, with coordinates given by
\begin{eqn}
\pi_{d,\text{red}}:~~~~~~~~~~~~~~~~~~\mu_d^{-1}(0)\subset M_d~~~~&\longrightarrow~~~{\tilde{M}_d}_0=\mu_d^{-1}(0)/\mF_d\\
(\v{X}_i,\v{P}_i,\v{A}_n,\v{Y}_n)~~~~&\longmapsto~~~(\v{X}_i,\v{V}_i,\v{B}_n,\v{E}_n),
\end{eqn}
where
\begin{eqn}
\v{X}_i&=\v{X}_i\\
\v{V}_i&=\v{P}_i-\v{A}(\v{X}_i)\\
\v{B}_n&=(\nabla_d^+\times\v{A})_n\\
\v{E}_n&=-\v{Y}_n.
\label{discRedHamilCoor}
\end{eqn}
As discussed earlier, care must be taken to ensure that the discrete fields $\v{B}_n$ and $\v{E}_n$ of ${\tilde{M}_d}_0$ obey the reduced manifold constraints
\begin{eqn}
(\nabla_d^+\cdot\v{B})_n&=0\\
-\nabla_d^-\cdot\sum\limits_{i=1}^L\int_{-\infty}^{\v{X}_i}\md\v{X}_i'W_{\sigma_1}(\v{X}_i'-\v{x}_n)-(\nabla_d^-\cdot\v{E})_n&=0.
\label{tildeMd0Constraints}
\end{eqn}

We note that these constraints must also be satisfied by any initial condition of the algorithm. The former condition is necessary to enforce a physically valid magnetic field. If the latter condition (Gauss's law) is not satisfied initially, it will have the effect of adding fixed `external' charges at the corresponding vertex $n$. In particular, a non-zero initial Gauss's law condition will evolve the system along some other reduced manifold ${{\tilde{M}_d}_\alpha=\mu_d^{-1}(\alpha)/\mF_d}$ with fixed external charge density $\alpha$.

A similar initial condition must be determined for the unreduced algorithm of Section~\ref{sectionCanonCons_PIC1} as well. The unreduced algorithm enforces the constraint ${(\nabla_d^+\cdot\nabla_d^+\times\v{A})_n=0}$ automatically. However, for simulations without external charges, care should be taken so that the value of ${(\mu_d(m))_n}$ in Eq.~(\ref{mu_d}) is everywhere initialized to zero. (Alternatively, Eq.~(\ref{mu_d}) can be used to properly initialize a simulation with external charges that remain fixed for all time.) We note that our derivation of the momentum map is essential to this correct specification of initial conditions.

We therefore proceed with the reduction of our discrete system and substitute Eq.~(\ref{discRedHamilCoor}) into the bracket of Eq.~(\ref{NewBracket}) to find
\begin{eqn}
\{\{F,G\}\}_d^{\text{red}}&[\v{X}_i,\v{V}_i,\v{B}_n,\v{E}_n]=\\
&\sum\limits_{i=1}^L\Bigg(\pd{F}{\v{X}_i}\cdot\pd{G}{\v{V}_i}-\pd{G}{\v{X}_i}\cdot\pd{F}{\v{V}_i}+\left[\pd{F}{\v{V}_i}\times\pd{G}{\v{V}_i}\right]\cdot\sum\limits_{n=1}^N\v{B}_nW_{\sigma_2}(\v{X}_i-\v{x}_n)\Bigg)\\
&+\sum\limits_{n=1}^N\Bigg(\left[\sum\limits_{i=1}^L\pd{F}{\v{V}_i}W_{\sigma_1}(\v{X}_i-\v{x}_n)-\left(\nabla_d^-\times\pd{F}{\v{B}}\right)_n\right]\cdot\pd{G}{\v{E}_n}\\
&\hspace{70pt}-\left[\sum\limits_{i=1}^L\pd{G}{\v{V}_i}W_{\sigma_1}(\v{X}_i-\v{x}_n)-\left(\nabla_d^-\times\pd{G}{\v{B}}\right)_n\right]\cdot\pd{F}{\v{E}_n}\Bigg).
\label{NewBracketRed}
\end{eqn}
To derive the ${\partial_{\v{V}_i}F\times\partial_{\v{V}_i}G}\cdot\v{B}(\v{X}_i)$ term in the bracket above, our interpolation functions were required to satisfy an additional constraint
\begin{eqn}
\nabla\times\sum\limits_{n=1}^N\v{A}_nW_{\sigma_1}(\v{x}-\v{x}_n)=\sum\limits_{n=1}^N(\nabla_d^+\times\v{A})_nW_{\sigma_2}(\v{x}-\v{x}_n)
\label{condOnW2}
\end{eqn}
for some interpolation function $W_{\sigma_2}$. As in Eq.~(\ref{condOnW}), this is a generalized, higher-dimensional analogue of the simplicial Whitney interpolant constraint.

Lastly, we re-express the Hamiltonian in the reduced coordinates of ${\tilde{M}_d}_0$ as
\begin{eqn}
H_d^{\text{red}}[\v{X}_i,\v{V}_i,\v{B}_n,\v{E}_n]=\frac{1}{2}\sum\limits_{i=1}^L\v{V}_i^2+\frac{1}{2}\sum\limits_{n=1}^N\Big(\v{E}_n^2+\v{B}_n^2\Big).
\end{eqn}
We have thus recovered the reduced Hamiltonian system of \citet{xiao_explicit_2015}.

As discussed in Section~\ref{subsectionReduceVM} for spatially continuous systems, this reduced Hamiltonian system is automatically guaranteed to preserve the momentum map of its parent, as long as it evolution is constrained to ${\tilde{M}_d}_0$. To see that this is the case, we may compute its evolution equations under the splitting scheme analogous to the unreduced case \citep{he_hamiltonian_2015,he_hamiltonian_2016,xiao_explicit_2015}
\begin{eqn}
H_d^{\text{red}}=\sum\limits_{\alpha=1}^3H_\v{V}^\alpha+H_\v{B}+H_\v{E}
\end{eqn}
where
\begin{eqn}
H_\v{V}^\alpha&\defeq\frac{1}{2}\sum\limits_{i=1}^L(V_i^\alpha)^2\\
H_\v{B}&\defeq\frac{1}{2}\sum\limits_{n=1}^N\v{B}_n^2\\
H_\v{E}&\defeq\frac{1}{2}\sum\limits_{n=1}^N\v{E}_n^2.
\end{eqn}
These subsystems generate the following EOM:
\begin{eqn}
H_\v{V}^\alpha
\begin{cases}
\dot{X}_i^\beta&=\delta_\alpha^\beta V_i^\alpha\\
\dot{V}_i^\beta&=\epsilon_{\beta\alpha\gamma}V_i^\alpha\sum\limits_{n=1}^NB_n^\gamma W_{\sigma_2}(\v{X}_i-\v{x}_n)\\
\dot{B}_n^\beta&=0\\
\dot{E}_n^\beta&=-\delta^\beta_\alpha\sum\limits_{i=1}^LV_i^\alpha W_{\sigma_1}(\v{X}_i-\v{x}_n)
\end{cases}
\label{KlimEOMRed}
\end{eqn}
\begin{eqn}
H_\v{B}
\begin{cases}
\dot{\v{X}}_i&=0\\
\dot{\v{V}}_i&=0\\
\dot{\v{B}}_n&=0\\
\dot{\v{E}}_n&=\left(\nabla_d^-\times\v{B}\right)_n
\end{cases}
\label{redEOM2}
\end{eqn}
\begin{eqn}
H_\v{E}
\begin{cases}
\dot{\v{X}}_i&=0\\
\dot{\v{V}}_i&=\sum\limits_{n=1}^N\v{E}_nW_{\sigma_1}(\v{X}_i-\v{x}_n)\\
\dot{\v{B}}_n&=-(\nabla_d^+\times\v{E})_n\\
\dot{\v{E}}_n&=0.
\end{cases}
\label{redEOM3}
\end{eqn}
We note again that $\alpha$ is not summed over in the expressions for subsystem $H_\v{V}^\alpha$.

Upon inspection, it is evident that the ${\tilde{M}_d}_0$ constraints of Eq.~(\ref{tildeMd0Constraints}) are obeyed in each subsystem when they are exactly solved. (As in the unreduced case, the above subsystems are readily exactly solved. In particular, note that ${\dot{V}_i^\alpha=0}$ in $H_\v{V}^\alpha$.) Consequently, the exact conservation law of the reduced system is systematically derived by simply expressing the unreduced momentum map of Eq.~(\ref{mu_d}) in ${\tilde{M}_d}_0$ coordinates
\begin{eqn}
(\bar{\mu}_d)_n&=-\nabla_d^-\cdot\sum\limits_{i=1}^L\int_{-\infty}^{\v{X}_i}\md\v{X}_i'W_{\sigma_1}(\v{X}_i'-\v{x}_n)-\nabla_d^-\cdot\v{E}_n\\
&\defeq\rho_n-\nabla_d^-\cdot\v{E}_n\\
&=0,
\label{reducedGaussLaw}
\end{eqn}
where in the final line we have noted that $\bar{\mu}_d$ vanishes by our previous choice of reduction to the preimage submanifold ${\mu_d}^{-1}(0)$.
Eq.~(\ref{reducedGaussLaw}) is Gauss's law, for which we are by construction guaranteed
\begin{eqn}
\dot{\bar{\mu}}_d=0,
\end{eqn}
as desired. (An analogous observation was made for the reduced bracket in \cite{kraus_gempic:_2017}, where the momentum map was treated as a Casimir.) The local charge conservation law of Eq.~(\ref{splitLocChargeCons})---whose expression is unmodified in the reduced submanifold---is furthermore satisfied, since ${\nabla_d^-\cdot\dot{\v{E}}_n=-\nabla_d^-\cdot\v{J}_n}$ holds in each subsystem of $H_d^{\text{red}}$.

\section{Conclusion\label{sectionConclusion}}

We have systematically derived conservation laws for both Lagrangian variational and Hamiltonian splitting PIC methods. Our approach for Lagrangian systems followed Noether's second theorem, while our approach for Hamiltonian systems employed the momentum map. Our treatment of Hamiltonian methods additionally revealed the decided advantage of gauge-compatible splitting methods over other time discretizations of Hamiltonian systems (see Section \ref{sectionMomMap}); when the sub-Hamiltonians of a splitting method are chosen to be gauge invariant and exactly solvable, such methods exactly preserve the momentum map associated with this gauge symmetry, as well as its corresponding conservation laws.

Our study further demonstrated the importance of deriving the momentum map of a discrete Hamiltonian system in order to correctly specify its initial conditions. In the case of gauge-invariant PIC methods, the momentum map's systematic definition of charge density (see Eq.~(\ref{chargeDefs})) enables the precise assignment (or, more commonly, avoidance) of `external' fixed charges at each lattice site $n$ as an initial condition.

The techniques we have developed are more widely applicable to the simulation of gauge theories, and in principle provide a template for the derivation of exact conservation laws in any gauge-symmetric variational or gauge-compatible splitting algorithm. Our classification of gauge-compatible splitting methods also provides a general framework for the construction of Hamiltonian splitting algorithms that obey exact conservation laws.

As a final note, these results convey an overall impression of the adaptability of gauge theories to the discrete structures of algorithms. Internal gauge symmetries are characterized by the transformation of fields defined against the background of space-time, and their geometric structure can therefore be maintained even after the algorithmic discretization of this background. The present effort demonstrates that much of the formalism that gauge theories employ in continuous space-time---whether in Lagrangian or Hamiltonian systems---is readily ported to discrete settings more suitable for computation.

\section{Acknowledgments}
We thank our anonymous reviewers for their helpful comments. This research was supported by the U.S. Department of Energy (DE-AC02-09CH11466). A.S.G. acknowledges the generous support of the Princeton University Charlotte Elizabeth Procter Fellowship.

% Create the reference section using BibTeX: %
\bibliographystyle{jpp}

\end{document}